\documentclass[aps,jcp,reprint,twocolumn,citeautoscript,longbibliography,preprintnumbers]{revtex4}
\usepackage{graphicx}
\usepackage{caption}
\usepackage{subcaption}
\usepackage{amsmath}
\usepackage{amssymb}
\usepackage{tabularx}
\usepackage{placeins}
\usepackage{units}
\usepackage{color}
\usepackage{hyperref}
\usepackage[normalem]{ulem}
\usepackage{dcolumn}
\usepackage{breakcites}
\usepackage{bm}
\usepackage{multirow}
\usepackage[export]{adjustbox}
\usepackage{mathtools}
\begin{document}

\preprint{{\large SAND2019-7914 O }}

\title{A new generation of effective core potentials from correlated calculations: 4s and 4p main group elements and first row additions}
\author{Guangming Wang$^{1,\dagger}$}
\email{gwang18@ncsu.edu}
\author{Abdulgani Annaberdiyev$^{1,\dagger}$, Cody A. Melton$^{1,2}$, M. Chandler Bennett$^{3}$, Luke Shulenburger$^2$, and Lubos Mitas$^1$}
\affiliation{
1) Department of Physics, North Carolina State University, Raleigh, North Carolina 27695-8202, USA \\
}
\affiliation{
2) Sandia National Laboratories, Albuquerque, New Mexico 87123, USA \\
}
\affiliation{{
3) Materials Science and Technology Division, Oak Ridge National Laboratory, \\ 
Oak Ridge, Tennessee, 37831, USA}}

\thanks{These authors contributed equally to this work}

\date{\today}
\begin{abstract}

Recently, we developed a new method for generating effective core potentials (ECPs) using valence energy isospectrality with explicitly correlated all-electron (AE) excitations and norm-conservation criteria.
We apply this methodology to the 3$^{rd}$-row main group elements, creating new correlation consistent effective core potentials
(ccECPs) and also derive additional ECPs to complete the ccECP table for H-Kr.
For K and Ca, we develop Ne-core ECPs and for the $4p$ main group elements, we construct [Ar]$3d^{10}$-core potentials.   
Scalar relativistic effects are included in their construction.
Our ccECPs reproduce AE spectra with significantly better accuracy than many existing pseudopotentials and show better overall consistency across multiple properties. 
The transferability of ccECPs is tested on monohydride and monoxide molecules over a range of molecular geometries. 
For the constructed ccECPs we also provide optimized DZ - 6Z valence Gaussian basis sets.

\end{abstract}

\maketitle

\section{Introduction}\label{intro}

Effective core potentials (ECPs) are widely used in electronic structure calculations of molecular and condensed systems. 
ECPs simplify these calculations by eliminating the core electrons, which usually do not significantly contribute to the valence electronic structure. Additionally, scalar and even spin-orbit relativistic effects can be included in the ECP in a  straightforward way. 
For some calculations, the use of ECPs is essentially unavoidable, especially for cases involving very heavy elements.
For instance, quantum Monte Carlo (QMC) methods scale poorly with nuclear charge, $Z$, as $\mathcal{O}(Z^{5.5-6.5})$ \cite{Ceperley1986,hrl87}; without the ability to reduce $Z$ via an ECP, QMC simulations would quickly become computationally prohibitive for heavy elements. 

In recent decades, a number of ECP tables have been developed from various construction schemes and concepts.
One such strategy is to use shape-consistency/norm-conservation of the single-particle orbitals, where for a given reference state the pseudo-orbitals are made to match the all-electron orbitals outside of a chosen cut-off radius.
An alternative approach has been to fit AE Hartree-Fock/Dirac-Fock (HF/DF) atomic excitations and gaps, known as energy consistency.
These published tables have led to substantial progress by enabling large-scale calculations of both molecular and condensed systems.
However, many of these studies have involved methods with mixed levels of accuracy such as Density Functional Theory (DFT) or correlated wave function approaches far from convergence.

Advancements in many-body theories and progress in computational platforms have made the quality of ECPs crucial for obtaining
high accuracy for larger systems and a broader range of elements. 
To the best of our knowledge, the use of many-body theories and incorporation of correlation effects in the development of ECPs has been rather limited \cite{Acioli, Maron, Fromager, TN-2013, TN-2015,eCEPP}.


Very recently, we have constructed correlation consistent ECPs (ccECPs) for the first two rows (period 2, 3) \cite{1-ccECP, 2-ccECP} as well as the $3d$ transition metal series \cite{3-ccECP}. 
In this work, we expand the ccECP library to include the 3rd-row main group elements, i.e, K, Ca, Ga, Ge, As, Se, Br, and Kr. 
We also derive additional ECPs, hitherto missing from the ccECP table, i.e. H, He, Li, Be, F, and Ne.

Let us briefly recall the ideas and the guidelines in the generation of ccECPs\cite{1-ccECP,2-ccECP,3-ccECP}:
\begin{enumerate}
    \item We use a many-body construction, i.e., the incorporation of correlation effects. 
    Fully correlated, relativistic AE calculations are taken as reference as opposed to mean-field frameworks such as HF or DFT.
    Coupled cluster with singles, doubles and perturbative triples (CCSD(T)) is used in the optimization.
    \item We use a simple and minimal parameterization for the ECP.
    The ccECPs are expressed in a well-established functional form \cite{BFD-2007} that is captured by a few gaussians per symmetry channel.
    \item We test the transferability of the ccECPs in molecular systems, ranging from compressed to stretched molecular geometries.
    This serves as a check on the ccECP outside equilibrium. 
    \item We establish an open database for future developments.
    The ccECPs are hosted at \url{http://pseudopotentiallibrary.org} with systematic labels and updates which is open to other contributors. 
\end{enumerate}

Obviously, our ccECPs do not represent the ultimate accuracy limit of pseudopotentials, although we believe that we have obtained 
results that are close to these accuracy limits 
within the given form, number of parameters, and
core choices considered.
Clearly, more elaborate parameterizations are possible for finer accuracy targets.
For instance, core-polarization effects and explicit spin-orbit coupling operators could be added subsequently, but that is beyond the scope of this paper.
We have tried to achieve chemical accuracy (1 kcal/mol) for all properties
and in many cases, we have been successful. 
In cases where this demand was too strict, we have constructed the ccECPs to show systematic and balanced errors that are more consistent than most of the published tables thus far.  

The paper is structured as follows.
We discuss the chosen functional form and parameterization for the ccECPs in Section \ref{ECP}.
In Section \ref{Objective}, the objective function and optimization strategies are examined.
The results for K and Ca are presented in Section \ref{sec:K,Ca} and Ga - Kr are presented in Section \ref{sec:Ga-Kr}.
We provide additional optimized ccECPs for selected elements in Section \ref{sec:additional} that includes H, He, Li, Be, F and Ne. 
Some of these additional ECPs also include an AE version that smooths out the Coulomb divergence in the potential. 
Together with our previous 
papers, this covers all the elements from H through Kr. 
The optimization of valence basis sets is discussed in Section \ref{sec:basis}.
Finally, we present conclusions and further discussions in Section \ref{Conclusions}.

\section{Methods}\label{methods}
\subsection{ECP Parametrization}\label{ECP}

The goal of this work is to approximate the relativistic AE Hamiltonian 
with a simpler valence only Hamiltonian, H$_{\rm val}$, where we assume the
Born-Oppenheimer approximation \cite{BO1927}.
The valence Hamiltonian is given by the following expression in atomic units (a.u):
\begin{equation}
   {\rm H}_{\rm val} = \sum_i[T^{\rm kin}_i + V^{\rm {ECP}}_i] +\sum_{i<j} 1/r_{ij}.
\end{equation}
We employ a semi-local ECP functional form given as:
\begin{equation}
    V^{\rm ECP}_i = V_{loc}(r_i) + \sum_{\ell=0}^{\ell_{max}} V_\ell(r_i) \sum_{m}|\ell m\rangle\langle \ell m|,
\end{equation}
where $i$ is the electron index and $r_i$ is the electron-ion distance.
The non-local term contains angular momentum projectors on $\ell m$-th state.
$\ell_{max}$ is the maximum angular momentum number of non-local channel projectors included in the ECP. 
Here, we choose $\ell_{max} = 2$ for $4p$ elements due to $3d^{10}$ electrons being in the core. 
For K and Ca, $\ell_{max} = 1$.
$V_{loc}(r_i)$ is the local channel where at the origin, Coulomb singularity is explicitly canceled out and $V_{loc}(r_i)$'s first derivative vanishes at the origin \cite{BFD-2007}:
\begin{equation}
    V_{loc}(r) = -\frac{Z_{\rm eff}}{r}(1 -e^{-\alpha r^2}) + \alpha Z_{\rm eff} re^{-\beta r^2} + \sum_{k=1}^{2} \gamma_{k} e^{-\delta_{k} r^2},
\end{equation} 
where $Z_{\rm eff}$ is the effective core charge, $Z_{\rm eff} = Z - Z_{\rm core}$ and $\alpha$, $\beta$, $\gamma_{k}$ and $\delta_{k}$ are coefficients to be optimized.
The choice of a non-singular form results in smooth orbitals that reduces computational resources such as  sizes of basis sets as
well as higher efficiency sampling in QMC.
The non-local term $V_\ell(r_i)$ has the following form:
\begin{equation}
    V_\ell(r) = \sum_{j=1}^{2}\beta_{\ell j} e^{-\alpha_{\ell j} r^2},
\end{equation}
where $\beta_{lj}$ are $ \alpha_{lj}$ are optimized non-local channel coefficients. 
This simple and compact form of ECP is chosen so that the ECPs can be readily used in standard quantum chemistry packages and other electronic structure codes in general. The simplicity of the form also makes the optimization process easier as commented upon later.


\subsection{Objective Function and Optimization Methodologies}\label{Objective}

Similarly to our previous ccECP constructions, the objective function is a weighted combination of several components such as energy consistency/isospectrality and terms related to orbital shape and norm conservation.
The energy consistency part of the objective function is given as follows
\begin{equation}
    \mathcal{E}^{2}=\sum_{s \in S}\left(\Delta E_{s}^{\mathrm{ECP}}-\Delta E_{s}^{\mathrm{AE}}\right)^{2},
\end{equation}
where the energy "gap" $\Delta E_{s}^{\mathrm{ECP}}$ is the CCSD(T) energy difference for the state $s$ relative to the ground state of the pseudoatom.
The states that we include are several neutral excitations, a number of ionized states, and electron affinity states for anions that are bounded. 
$\Delta E_{s}^{\mathrm{AE}}$ is the corresponding quantity in the all-electron atom with 10$^{th}$ order scalar relativistic Douglass-Kroll-Hess Hamiltonian \cite{DKH}.
Therefore, relativistic effects are inherently included in our ccECPs.

It is important to note that all AE calculations in this work are fully correlated.
In other words, they include valence-valence (VV), core-valence (CV) and core-core correlations (CC).
Typically, ECPs include only VV correlations, however, the inclusion of  CC+CV+VV correlations in the AE reference is essential for several aspects as well as in the optimization as it will be shown in  Section \ref{Results}. 

Isospectrality is a crucial property of an ECP since it relies on actual observable quantities. 
However, as we have shown in  previously, the optimization of the spectrum alone can result in reduced transferability of the ECP
since  the atomic spectrum can be overfit at the expense of other properties.
Therefore, we make use of additional norm/shape conservation related function which is given as:
\begin{equation}
\begin{aligned} 
\mathcal{N}^{2}=& \sum_{\ell} \left(\epsilon_{\ell}^{\mathrm{ECP}}-\epsilon_{\ell}^{\mathrm{AE}}\right)^{2}+\gamma \Big[\left(N_{\ell}^{\mathrm{ECP}}-N_{\ell}^{\mathrm{AE}}\right)^{2} \\ &+\left(V_{\ell}^{\mathrm{ECP}}-V_{\ell}^{\mathrm{AE}}\right)^{2}+\left(S_{\ell}^{\mathrm{ECP}}-S_{\ell}^{\mathrm{AE}}\right)^{2}\Big],
\end{aligned}
\label{eqn:norm}
\end{equation}
where the following ECP quantities are matched to relativistic AE counterparts as explained below.
$\epsilon_{\ell}^{\mathrm{AE}}$ are one-particle eigenvalues and these are matched for all $\ell$ angular momentum channels. 
To evaluate the quantities inside the brackets, a cut-off radius $R_{\ell}$ is chosen as the outer-most extremum of $r^{\frac{4}{5}} \phi_{\ell}^{\mathrm{AE}}$.
Then, the norm $N_{\ell}^{\mathrm{AE}}=\int_{0}^{R_{\ell}}\left(r^{\ell+1} \phi_{\ell}^{\mathrm{AE}}(r)\right)^{2} \mathrm{d}r$ 
within that radius, the value $V_{\ell}^{\mathrm{AE}}=\phi_{\ell}^{\mathrm{AE}}\left(R_{\ell}\right)$ 
and derivative/slope $S_{\ell}^{\mathrm{AE}}=\frac{\mathrm{d}}{\mathrm{d} r} \phi_{\ell}^{\mathrm{AE}}\left.(r)\right|_{R_{\ell}}$ at that radius are matched.
Usually, setting $\gamma=0$ resulted in accurate molecular properties, however, in some cases, we set $\gamma=1$ to further improve the transferability.
The total objective function is then:
\begin{equation}
    O^{2}=w_{0}
    \mathcal{E}^{2}+\mathcal{N}^{2},
\end{equation}
where $w_{0}$ was chosen to be 0.05, although adjustments were made in some cases. 

Calculation of $\mathcal{E}^{2}$ is the most expensive/demanding part of the objective function evaluation since 
 we included $\sim 10$ atomic state energies where each must be evaluated at CCSD(T) level with a large, uncontracted basis set. 
In order to alleviate this cost, we proposed a recipe to significantly speed up the optimization procedure in our $3d$ transition metal paper \cite{3-ccECP}.
Here, we briefly recall that approach.
It is known that the correlation energies of various ECPs, even constructed in different approaches are essentially the same, unlike their corresponding HF energies.
This allows one to map the CCSD(T) gaps to a mean-field gap by shifting the corresponding HF energy with a quantity related to differences in AE and ECP correlation energies.
Therefore, the ECP correlation energies need not be calculated at every step of the optimization and the objective function is evaluated with a cheaper, appropriately shifted HF calculation. 
 Furthermore, this enables us to use relatively small basis-sets (unc-aug-ccpwCTZ) in the optimization since the HF-level
 complete basis-set (CBS) values are reached quickly.
This methodology allows us to use CCSD(T) in the optimization in a computationally less costly manner while keeping essentially the same level of accuracy.
More detailed discussion is given in the aforementioned work.

The quantum chemistry packages \textsc{Molpro}\cite{MOLPRO} and \textsc{PySCF}\cite{PYSCF} were used for evaluation of the objective functions throughout this work.
The nonlinear optimization code \textsc{Donlp2}\cite{donlp2} was used to optimize the ccECP parameters. 
Most ccECP parameters in the optimization are initialized either from BFD\cite{BFD-2007} or STU\cite{STU-SDF,STU-MWB,STU-MDF} pseudopotentials.


\section{Results}\label{Results}

In this section, we provide the atomic spectrum results and molecular transferability tests for all the elements mentioned above. 
For each element, we show the quality of the spectrum by tabulating the mean absolute deviation (MAD) of the ECP gaps with respect to AE values.
The MAD is given as:
\begin{equation}
    \mathrm{MAD}=\frac{1}{N_{s}} \sum_{s=1}^{N_{s}}\left|\Delta E_{s}^{\mathrm{ECP}}-\Delta E_{s}^{\mathrm{AE}}\right| .
\end{equation}
For Ga - Kr these quantities and errors are provided at CBS limit and for the other lighter atoms these are given at unc-aug-cc-pCV5Z basis set.
In order to obtain the CBS limit energy, SCF and correlation energies are extrapolated separately.
The HF energy extrapolation has the following form
\begin{equation}
    E^{\rm HF}_n = E^{\rm HF}_{\rm CBS} + a\exp(-bn),
    \label{eqn:hfextrap}
\end{equation}
where $n$ is the basis size, $E_{\rm CBS}^{\rm HF}$ is the HF CBS limit value, and $a$ and $b$ are fit parameters. 
The correlation energy is extrapolated as
\begin{equation}
    E^{\rm corr}_n = E^{\rm corr}_{\rm CBS} + \frac{{c}}{(n+3/8)^{3}} + \frac{{d}}{(n+3/8)^5},
    \label{eqn:corrextrap}
\end{equation}
where $c$ and $d$ are other fit parameters.
The basis set used in extrapolation is uncontracted aug-cc-pwCVnZ\cite{K-Ca-basis,basis-4p} where n=(T, Q, 5).

We also evaluated the errors of other established ECPs so as to demonstrate the degree of improvement in our ccECP pseudopotentials.
The ECPs considered are BFD\cite{BFD-2007}, YSTU (Y=\{SDF\cite{STU-SDF}, MWB\cite{STU-MWB}, MDF\cite{STU-MDF}\}), SBKJC\cite{SBKJC-12,SBKJC-345}, eCEPP\cite{eCEPP}, CRENBL\cite{CRENBL-12,CRENBL-3}. 
We also present the calculations for an AE uncorrelated-core (UC), where there are only VV correlations considered and the core is of the same size as in the ECPs.
The same uncontracted basis sets and extrapolation schemes are used for these core approximations for consistency.
We note that
 only VV correlations are present in our ccECP calculations.
 In addition, we want to emphasize that we did not include core polarization/relaxation (CPP) terms into ccECPs, neither considered these when using the other existing tables. One reason for this choice has been our primary goal to probe for the limits of accuracy of the very simplest ECP form. Additional reason has been our aim for the widest possible use and the fact that 
  in majority of condensed/periodic matter codes such terms are not currently implemented. Clearly, this can be addressed in future studies.

For each element, the quality/transferability of the ccECP is tested by comparing the binding energy of the ECP for selected molecules to the fully correlated AE case which utilizes relativistic DKH Hamiltonian. 
The binding energy is calculated for a variety of molecular geometries, including for compressed geometries up to the dissociation
bond length, having in mind potential high-pressure applications.
The discrepancy $\Delta(r)$ between the ECP and AE is plotted for all calculated molecules:
\begin{equation}
    \Delta(r)=D_e(r)^{\rm ECP} - D_e(r)^{\rm AE}_{\rm DKH}.
\end{equation}
We fit the calculated binding energies from multiple bond lengths to Morse potential:
\begin{equation}
\label{morse_pot}
    V(r)=D_{e}\left(e^{-2 a\left(r-r_{e}\right)}-2 e^{-a\left(r-r_{e}\right)}\right),
\end{equation}
where $D_{e}$ is the dissociation energy, $r_e$ is the equilibrium bond length and the parameter, $a$, is used to calculate the vibrational frequency
\begin{equation}
    \omega_{e}=\sqrt{\frac{2 a^{2} D_{e}}{\mu}},
\end{equation}
where $\mu$ is the reduced mass of the molecule.
Fitted $D_e$, $r_e$, $\omega_e$ and $D_{diss}$ (binding energy at the dissociation threshold at the short bond lengths) are compared to AE counterparts to assess the transferability of an ECP for a particular molecule.

We consider the monohydride and monoxide molecules as our test systems to probe the $\sigma$ and $\pi$ bonds. 
In some cases, the homonuclear dimer was also used if it formed the same type of bonds as in the above cases.
Other ECP tests are included as well for comparison.
Each oxygen ECP is taken from the corresponding ECP table and our ccECP hydrogen is used throughout all hydride tests.

All transferability tests were carried out at unc-aug-cc-pwCV5Z basis sets unless otherwise specified.
From limited tests on SeO, we observed that no significant changes occur in the discrepancy of binding energies $\Delta(r)$ compared to the CBS limit (see supplementary material).

\begin{figure}[htbp!]
\caption{MAD and LMAD for the $3^{rd}$-row main group elements. 
For each atom, we provide MAD of the entire calculated spectrum and MAD of selected low-lying atomic states (LMAD). 
The LMAD atomic states include electron affinity, first and second ionizations only.
The solid lines represent MAD and the dashed lines represent LMAD.
The shaded region indicates the band of chemical accuracy.}
\label{fig:mad_graph}
\includegraphics[width=\columnwidth]{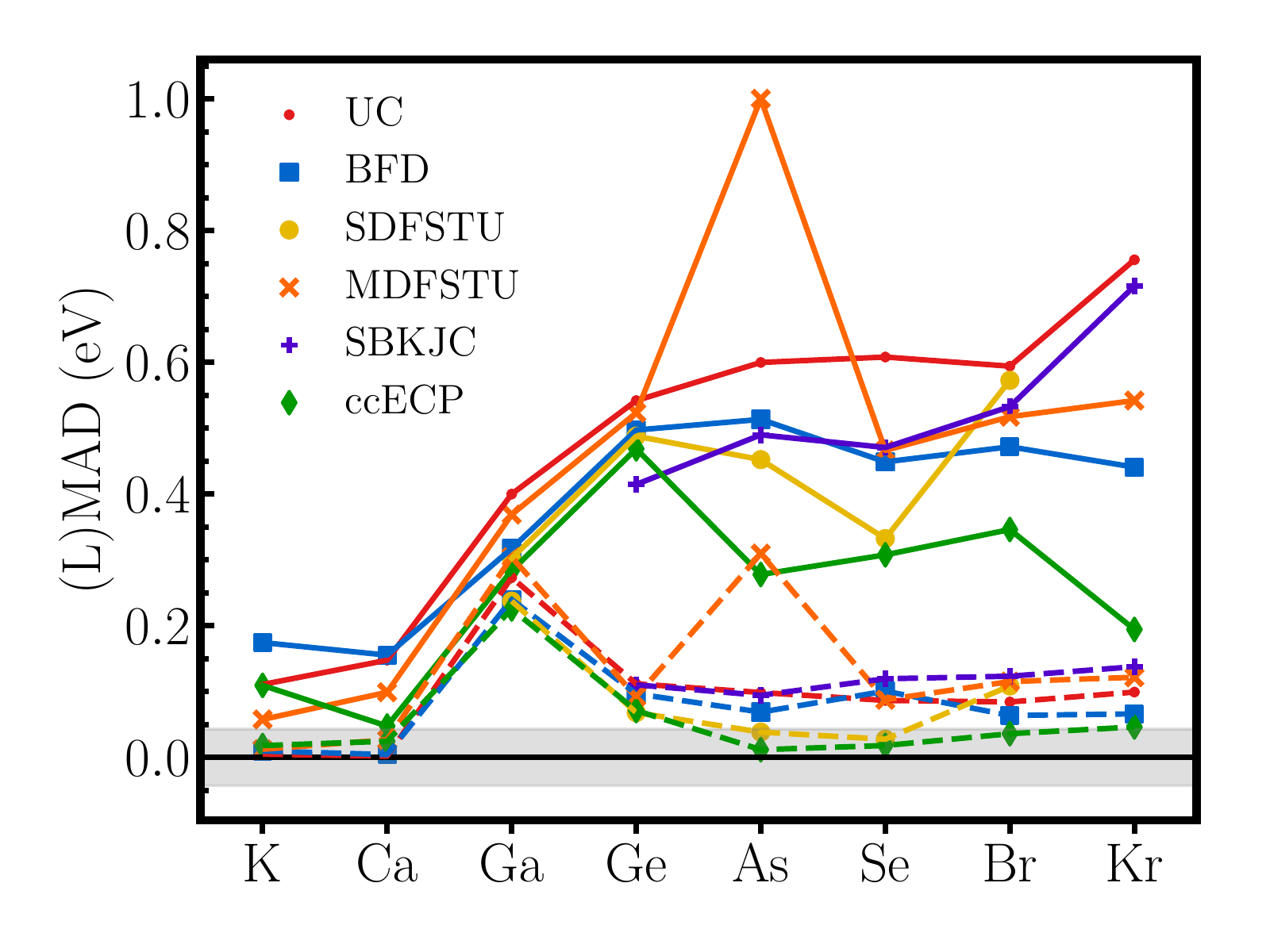}
\centering
\end{figure}

In table \ref{tab:MADs}, we show a summary of spectrum discrepancies of all examined ECPs compared to AE spectrum. 
The MAD of the entire spectrum and the MAD of some selected low-lying states (LMAD) are given while for
Li and Be we give the MAD values only. 
MDFSTU and MWBSTU are combined in one column since they are mutually exclusive for each of the atoms considered in this work.
Only available ECP MADs are listed with the rest left blank. 
The calculated AE spectrum and ECP discrepancies for each atom can be found in the supplementary material. 
A summary graph of MADs and LMADs for $3^{rd}$-row main group elements is shown in figure \ref{fig:mad_graph}.

All optimized ccECP parameters are provided in tables \ref{tab:K-Kr_params} and \ref{tab:additional_params} in semi-local form.
The tables assume the following common format of the potential
\begin{equation}
    V_{\ell}(r)=\sum_{k} \beta_{\ell k} r^{n_{\ell k}-2} e^{-\alpha_{\ell k} r^{2}},
\end{equation}
where $\ell$ is the angular momentum channel.
Note that $-\frac{Z_{\rm eff}}{r}$ term is not tabulated since it is added by default in most codes/packages.

Detailed discussions about each element and the accuracy of the ccECPs is given in Sections \ref{sec:K,Ca},\ref{sec:Ga-Kr},\ref{sec:additional}. 
The core radii of ccECPs are tabulated in table \ref{tab:core_radii} and valence basis sets are discussed in Section \ref{sec:basis}.

\begin{table*}[htbp!]
\setlength{\tabcolsep}{4pt} 
\small
\centering
\caption{ Parameter values for 3$^{rd}$-row main group element ccECPs. For all ECPs, the highest $\ell$ value corresponds to the local channel $\ell_{loc}$.
Note that the highest non-local angular momentum channel $\ell_{max}$ is related to it as $\ell_{max}=\ell_{loc}-1$.
}
\label{tab:K-Kr_params}
\begin{tabular}{ccrrrrrccrrrrr}
\hline\hline
\multicolumn{1}{c}{Atom} & \multicolumn{1}{c}{$Z_{\rm eff}$} & \multicolumn{1}{c}{$\ell$} & \multicolumn{1}{c}{$n_{\ell k}$} & \multicolumn{1}{c}{$\alpha_{\ell k}$} & \multicolumn{1}{c}{$\beta_{\ell k}$} & & \multicolumn{1}{c}{Atom} & \multicolumn{1}{c}{$Z_{\rm eff}$} & \multicolumn{1}{c}{$\ell$} & \multicolumn{1}{c}{$n_{\ell k}$} & \multicolumn{1}{c}{$\alpha_{\ell k}$} & \multicolumn{1}{c}{$\beta_{\ell k}$} \\
\hline

K  &  9 &  0 &  2 & 11.442508 &  11.866873  && As &  5 &  0 &  2 & 3.479387 &  75.655194       \\
   &    &  0 &  2 &  6.537124 &  90.076771  &&    &    &  0 &  2 & 1.637480 &  -3.311453       \\
   &    &  1 &  2 &  9.631219 &  11.534202  &&    &    &  1 &  2 & 3.229364 &  67.961867       \\
   &    &  1 &  2 &  4.508811 &  27.720235  &&    &    &  1 &  2 & 1.666366 &  -3.094558       \\
   &    &  2 &  1 &  7.273863 &   9.000000  &&    &    &  2 &  2 & 2.068163 &  24.304734       \\
   &    &  2 &  3 & 11.172983 &  65.464770  &&    &    &  2 &  2 & 1.546999 &   0.939456       \\
   &    &  2 &  2 &  7.706175 & -10.844336  &&    &    &  3 &  1 & 1.285931 &   5.000000       \\
   &    &  2 &  2 &  5.624917 & -15.963161  &&    &    &  3 &  3 & 9.934874 &   6.429657       \\
   &&&&&                                    &&    &    &  3 &  2 & 1.895682 & -15.012439       \\
   &&&&&                                    &&    &    &  3 &  2 & 1.728256 &   2.898814       \\
   &&&&&                                                                                       \\
Ca & 10 &  0 &  2 & 11.240167 & 149.302623  && Se &  6 &  0 &  2 & 4.175363 &  71.379280       \\
   &    &  0 &  2 &  5.353612 &  23.759329  &&    &    &  0 &  2 & 2.144911 &   0.426199       \\
   &    &  1 &  2 & 13.066548 &  99.204114  &&    &    &  1 &  2 & 4.287722 &  50.948290       \\
   &    &  1 &  2 &  4.027485 &  13.452161  &&    &    &  1 &  2 & 2.095383 &   5.542881       \\
   &    &  2 &  1 &  7.041332 &  10.000000  &&    &    &  2 &  2 & 1.394037 &   6.204697       \\
   &    &  2 &  3 & 14.014449 &  70.413317  &&    &    &  2 &  2 & 1.696599 &   0.533957       \\
   &    &  2 &  2 & 13.769362 & -92.872980  &&    &    &  3 &  1 & 2.977052 &   6.000000       \\
   &    &  2 &  2 &  4.717260 &  -5.753568  &&    &    &  3 &  3 & 7.016674 &  17.862311       \\
   &&&&&                                    &&    &    &  3 &  2 & 3.960663 & -20.009132       \\
   &&&&&                                    &&    &    &  3 &  2 & 5.028263 &  10.005735       \\
   &&&&&                                                                                       \\
 Ga &  3 &  0 &  2 &  1.857811 &  21.789310 && Br &  7 &  0 &  2 & 4.971807 &  85.884347 \\
    &    &  0 &  2 &  0.919506 &  -2.866851 &&    &    &  0 &  2 & 2.042687 &   4.621255 \\
    &    &  1 &  2 &  1.920302 &  18.639860 &&    &    &  1 &  2 & 4.711839 &  55.361715 \\
    &    &  1 &  2 &  1.008959 &  -1.633697 &&    &    &  1 &  2 & 2.384293 &  11.031410 \\
    &    &  2 &  2 &  0.627509 &   2.035237 &&    &    &  2 &  2 & 3.412863 &  26.410410 \\
    &    &  2 &  2 &  0.326190 &  -0.085324 &&    &    &  2 &  2 & 1.530285 &   5.468739 \\
    &    &  3 &  1 & 17.004739 &   3.000000 &&    &    &  3 &  1 & 3.665770 &   7.000000 \\ 
    &    &  3 &  3 & 14.999618 &  51.014218 &&    &    &  3 &  3 & 5.293023 &  25.660393 \\
    &    &  3 &  2 & 11.992792 & -39.000626 &&    &    &  3 &  2 & 3.176376 &  13.040262 \\
    &    &  3 &  2 & 14.992823 &  35.446594 &&    &    &  3 &  2 & 2.897544 & -21.908839 \\
    &&&&&                                                                                \\
 Ge &  4 &  0 &  2 & 2.894474 &  43.265429  && Kr &  8 &  0 &  2 &  5.490729 &  92.889552 \\
    &    &  0 &  2 & 1.550340 &  -1.909340  &&    &    &  0 &  2 &  3.863012 &  12.929478 \\
    &    &  1 &  2 & 2.986529 &  35.263014  &&    &    &  1 &  2 &  4.038577 &  43.099524 \\
    &    &  1 &  2 & 1.283381 &   0.963440  &&    &    &  1 &  2 &  3.306789 &   9.509760 \\
    &    &  2 &  2 & 1.043001 &   2.339019  &&    &    &  2 &  2 &  4.213480 &  17.804945 \\
    &    &  2 &  2 & 0.554563 &   0.541381  &&    &    &  2 &  2 &  1.549897 &   4.589115 \\
    &    &  3 &  1 & 1.478963 &   4.000000  &&    &    &  3 &  1 & 10.794238 &   8.000000 \\
    &    &  3 &  3 & 3.188906 &   5.915851  &&    &    &  3 &  3 & 13.323389 &  86.353904 \\
    &    &  3 &  2 & 1.927439 & -12.033713  &&    &    &  3 &  2 &  9.292050 & -11.114533 \\
    &    &  3 &  2 & 1.545539 &   1.283543  &&    &    &  3 &  2 & 20.148958 &  10.229519 \\
   &&&&&                                                                                  \\                                    
\hline\hline
\end{tabular}
\end{table*}

\begin{table*}
\centering
\caption{A summary of the spectrum discrepancies of various ECPs for the 3$^{rd}$-row main group elements. For each atom, we provide MAD of the entire calculated spectrum and MAD of selected low-lying atomic states (LMAD). The LMAD atomic states include electron affinity, first and second ionizations only. All values are in eV.}
\label{tab:MADs}
\begin{tabular}{lcccccccccccccc}
\hline\hline
Atom &  Quantity   &   UC &    BFD &    SDFSTU   & MDFSTU/MWBSTU  &     SBKJC &  CRENBL   &  ccECP  \\
\hline
K         & MAD &  0.110808 &  0.174431 &            &  0.057621  &           & 0.123791 &  0.109417 \\
          &LMAD &  0.004905 &  0.009807 &            &  0.012360  &           & 0.015636 &  0.017944 \\
\hline   
Ca        & MAD &  0.147679 &  0.155381 &            &  0.098974  &           & 0.199446 &  0.047605 \\
          &LMAD &  0.001868 &  0.004282 &            &  0.026807  &           & 0.002046 &  0.024634 \\
\hline   
Ga        & MAD &  0.400024 &  0.317379 &  0.302297  &  0.368861  &           &          &  0.283783 \\
          &LMAD &  0.273255 &  0.239687 &  0.237425  &  0.304068  &           &          &  0.225519 \\
\hline   
Ge        & MAD &  0.542077 &  0.497299 &  0.487997  &  0.523465  &  0.414607 &          &  0.468781 \\
          &LMAD &  0.111700 &  0.096646 &  0.068051  &  0.092133  &  0.110203 &          &  0.071163 \\
\hline   
As        & MAD &  0.599997 &  0.513494 &  0.452386  &  1.000536  &  0.490010 &          &  0.277827 \\
          &LMAD &  0.098619 &  0.068839 &  0.038426  &  0.309579  &  0.094349 &          &  0.011692 \\
\hline   
Se        & MAD &  0.608203 &  0.448970 &  0.332396  &  0.466401  &  0.470802 &          &  0.307693 \\
          &LMAD &  0.086533 &  0.100668 &  0.027611  &  0.087471  &  0.119486 &          &  0.018102 \\
\hline   
Br        & MAD &  0.594315 &  0.471870 &  0.572803  &  0.517565  &  0.532889 &          &  0.346153 \\
          &LMAD &  0.084249 &  0.063693 &  0.108582  &  0.115210  &  0.123326 &          &  0.036020 \\
\hline   
Kr         & MAD &  0.755949 &  0.440971 &           &  0.542141  &  0.716075 &          &  0.194395 \\
           &LMAD &  0.099244 &  0.066169 &           &  0.121735  &  0.137783 &          &  0.046252 \\
\hline
\hline
\end{tabular}
\end{table*}

\subsection{K and Ca atoms}\label{sec:K,Ca}

In this subsection, we explain the results for K and Ca atoms. 
These atoms use Ne-core ECPs and they involve only $s$ and $p$ non-local projectors.

\begin{figure*}[!htbp]
\centering
\begin{subfigure}{0.5\textwidth}
\includegraphics[width=\textwidth]{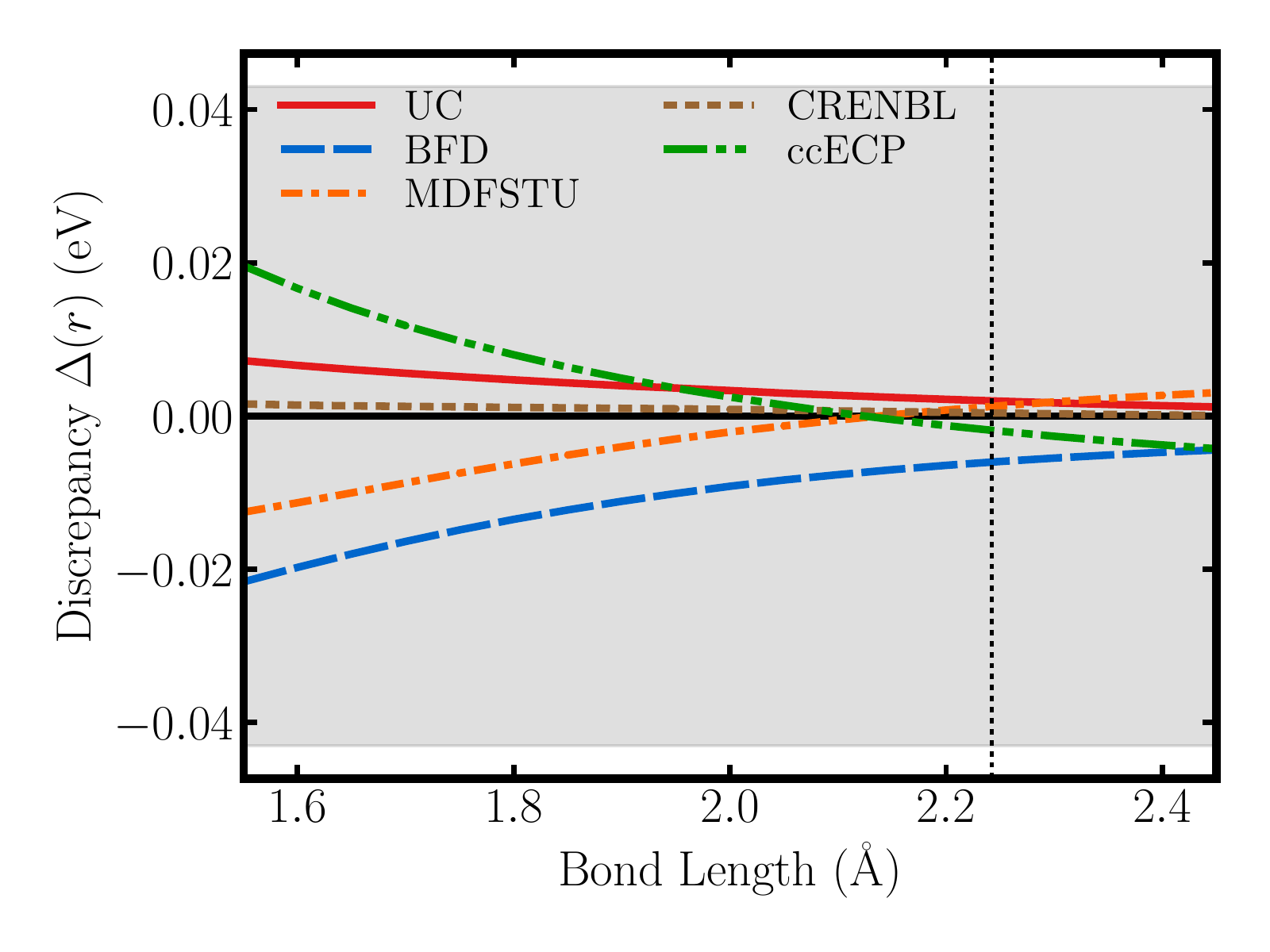}
\caption{KH binding curve discrepancies}
\label{fig:KH}
\end{subfigure}%
\begin{subfigure}{0.5\textwidth}
\includegraphics[width=\textwidth]{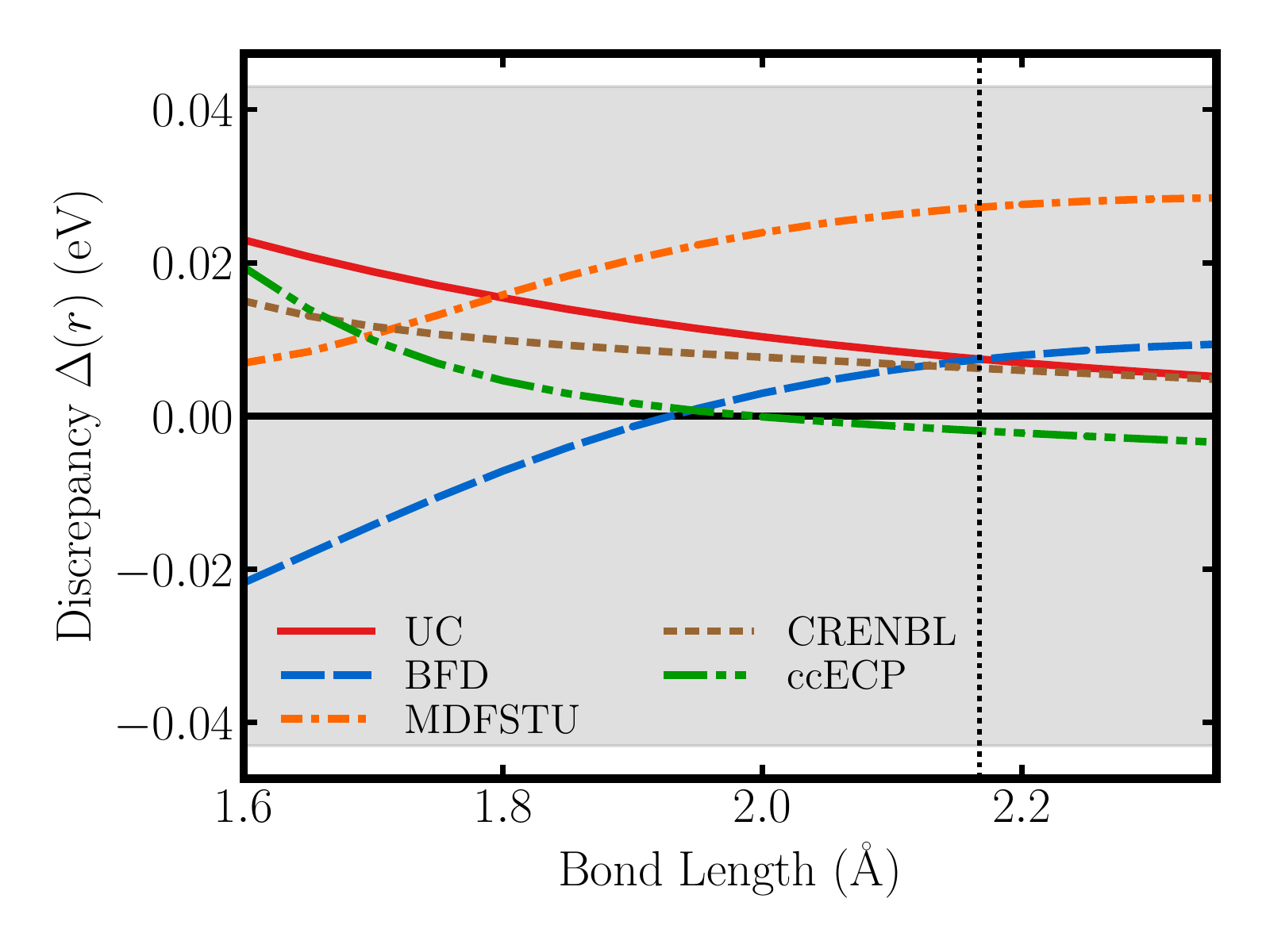}
\caption{KO binding curve discrepancies}
\label{fig:KO}
\end{subfigure}
\caption{Binding energy discrepancies for (a) KH and (b) KO molecules with regard to CCSD(T). The shaded region indicates the band of chemical accuracy. The dashed vertical line represents the equilibrium geometry.}
\label{fig:K_mols}
\end{figure*}

\begin{table}[!htbp]
\centering
\caption{Mean absolute deviations of binding parameters for various core approximations with respect to AE data for K, Ca hydride and oxide molecules. All parameters were obtained using Morse potential fit. The parameters shown are dissociation energy $D_e$, equilibrium bond length $r_e$, vibrational frequency $\omega_e$ and binding energy discrepancy at dissociation bond length $D_{diss}$.}
\label{morse:K,Ca}
\begin{tabular}{l|rrrrrrrrrr}
\hline\hline
{} & $D_e$(eV) & $r_e$(\AA) & $\omega_e$(cm$^{-1}$) & $D_{diss}$(eV) \\
\hline
UC         &  0.007(2) &   0.002(1) &  1.2(2.3)  &        0.02(3) \\
BFD        &  0.021(2) &   0.003(1) &  2.8(2.3)  &        0.05(3) \\
CRENBL     &  0.007(2) &   0.001(1) &  0.6(2.3)  &        0.02(3) \\
MDFSTU     &  0.030(2) &   0.002(1) &  2.6(2.3)  &        0.03(3) \\
ccECP      &  0.007(2) &   0.001(1) &  1.9(2.3)  &        0.02(3) \\
\hline\hline
\end{tabular}
\end{table}

Table \ref{tab:MADs} shows the MAD and LMAD of the spectrum discrepancies of various core approximations for the K atom.
The chosen spectrum includes neutral $d$ excitations, electron affinity state, and all ionizations until [Ne]$3s^2$.
Here, the least MAD corresponds to MDFSTU and the least LMAD corresponds to BFD. 
Figure \ref{fig:K_mols} shows binding energy discrepancy $\Delta(r)$ of different ECPs for KH and KO molecules.
The shaded region indicates the chemical accuracy of 0.043 eV and the vertical dashed line marks the equilibrium bond length.
Although all ECPs are within chemical accuracy, the ccECP is closest to zero and flatter near the KO molecule equilibrium geometry.  
Note that a flat discrepancy curve will result in a better vibrational frequency from the fit.
It should be noted that CRENBL ECP produces also very accurate molecular properties for KH and KO, however, it has singularities 
in both the attractive/local and repulsive nonlocal channels.
Our ccECP has a reasonable balance of MAD and LMAD but arguably is more accurate in transferability, especially near the equilibrium region.

\begin{figure*}[!htbp]
\centering
\begin{subfigure}{0.5\textwidth}
\includegraphics[width=\textwidth]{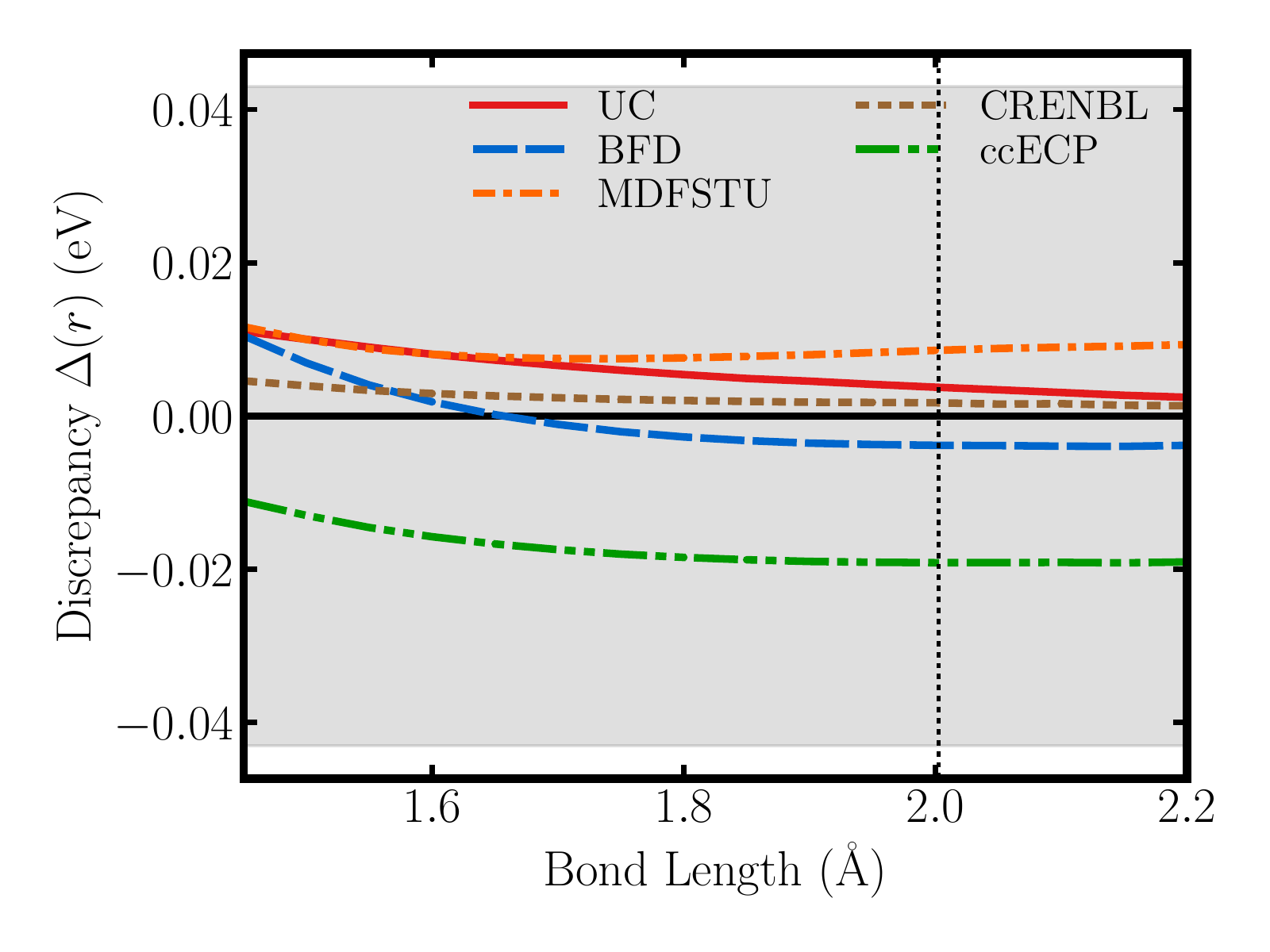}
\caption{CaH binding curve discrepancies}
\label{fig:CaH}
\end{subfigure}%
\begin{subfigure}{0.5\textwidth}
\includegraphics[width=\textwidth]{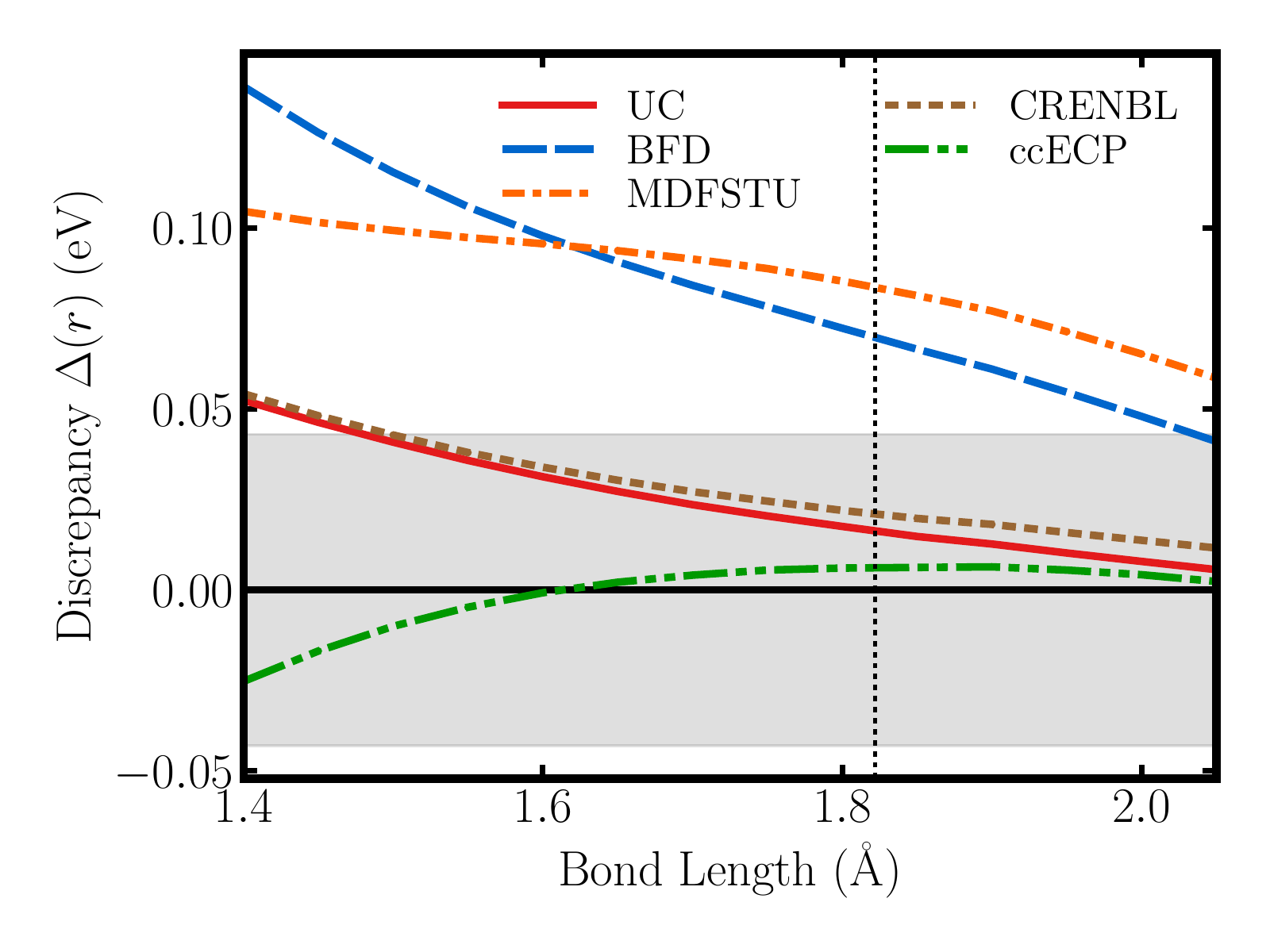}
\caption{CaO binding curve discrepancies}
\label{fig:CaO}
\end{subfigure}
\caption{Binding energy discrepancies for (a) CaH and (b) CaO molecules. The rest of of the notation is the same as for KH and KO systems.}
\label{fig:Ca_mols}
\end{figure*}

Atomic data for Ca atom is given in table \ref{tab:MADs} as well.
Here, ccECP has a significantly lower MAD in the spectrum compared to all other core approximations.
The lowest LMAD corresponds to CRENBL although its MAD is still significantly higher than ccECP. 
Binding energy discrepancies for CaH and CaO are shown in figure \ref{fig:Ca_mols}.
We see that in CaO, most core approximations underbind as the bond is compressed whereas ccECP has minimal errors throughout all geometries.

In table \ref{morse:K,Ca}, we give the MAD of all fitted parameters of Morse potential in Eqn. (\ref{morse_pot}) for K and Ca molecules (KH, KO, CaH, CaO) as a summary of transferability.
We find that the ccECP results in almost the same transferability as CRENBL with a minor exception in $\omega_e$. 
It is again worthwhile to mention that CRENBL, SBKJC, and STU ECP parameterizations differ from BFD, eCEPP, and ccECP form where the Coulomb and other singularities 
are absent by construction.
Therefore, we conjecture that the accuracy achieved in ccECP is close to the best possible fit within the given form and additional optimized gaussians will likely result in only mild and diminishing improvements.   

In summary, we see that due to the valence space being large in K and Ca, chemical accuracy can be achieved for molecules and for a large part of the atomic spectrum. 

\subsection{Ga - Kr atoms}\label{sec:Ga-Kr}

This subsection discusses the results for $4p$ elements: Ga - Kr.
These atoms use a large [Ar]$3d^{10}$ core ECPs with $s$, $p$, and $d$ non-local channel operators.
In this set of atoms, the valence space(4s and 4p) is smaller and the energy scales are significantly higher, resulting in larger errors in both spectrum and molecular binding energies.
This is especially pronounced for higher ionization states and in compressed bond regions. 

\begin{figure*}[!htbp]
\centering
\begin{subfigure}{0.5\textwidth}
\includegraphics[width=\textwidth]{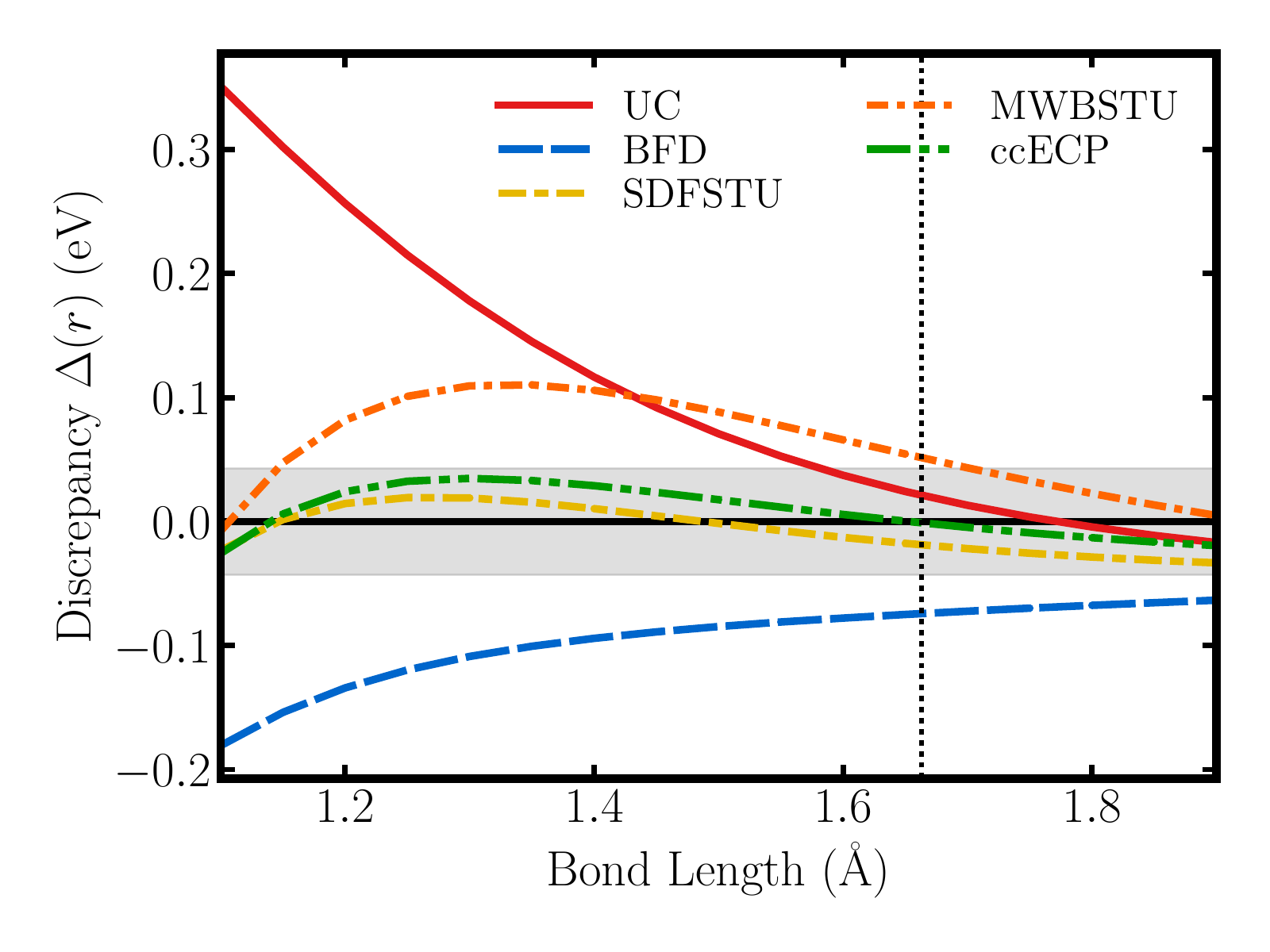}
\caption{GaH binding curve discrepancies}
\label{fig:GaH}
\end{subfigure}%
\begin{subfigure}{0.5\textwidth}
\includegraphics[width=\textwidth]{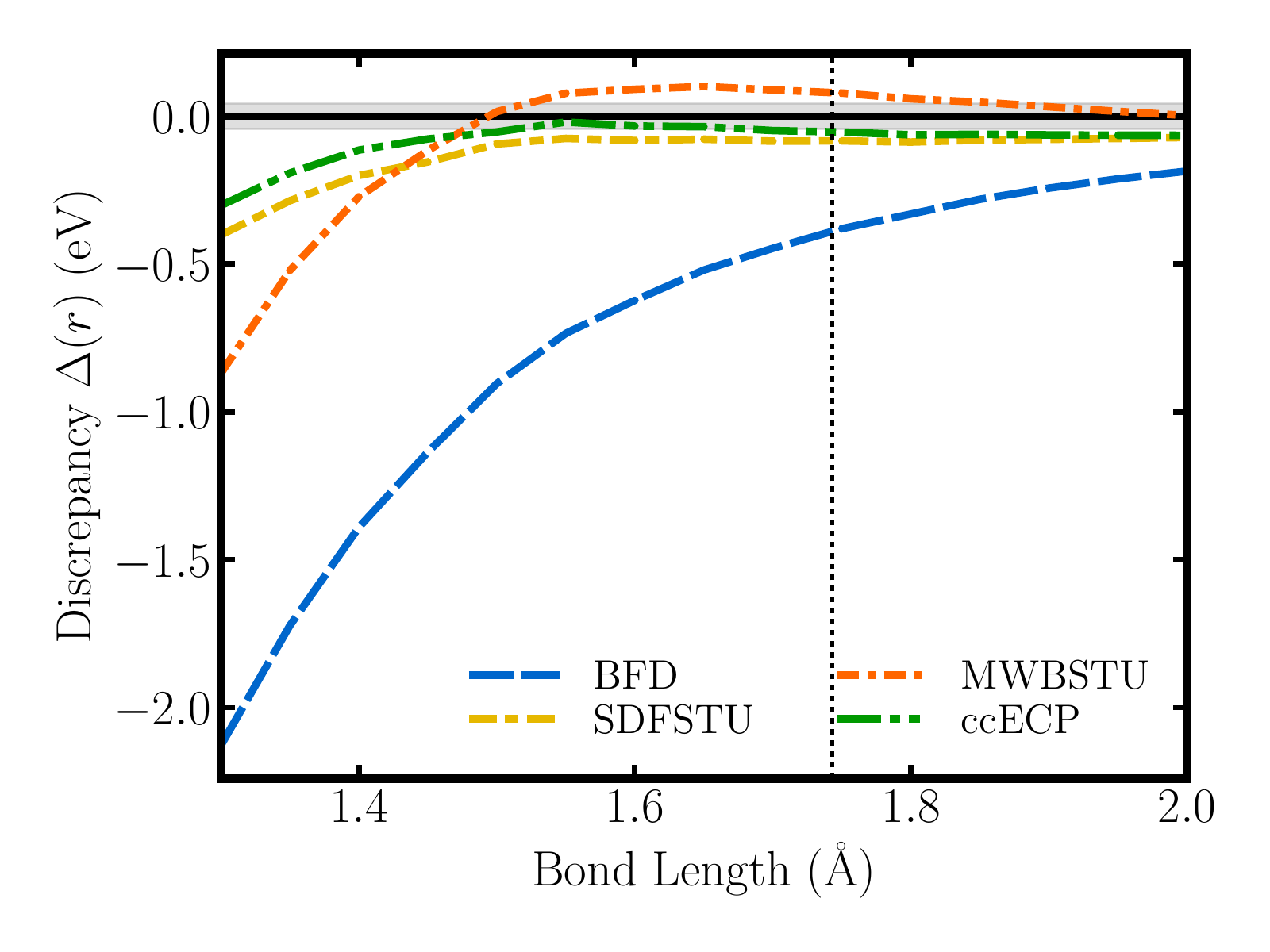}
\caption{GaO binding curve discrepancies}
\label{fig:GaO}
\end{subfigure}
\caption{Binding energy discrepancies for (a) GaH and (b) GaO molecules.
The same notation applies as for the previous cases.}
\label{fig:Ga_mols}
\end{figure*}

\begin{figure*}[!htbp]
\centering
\begin{subfigure}{0.5\textwidth}
\includegraphics[width=\textwidth]{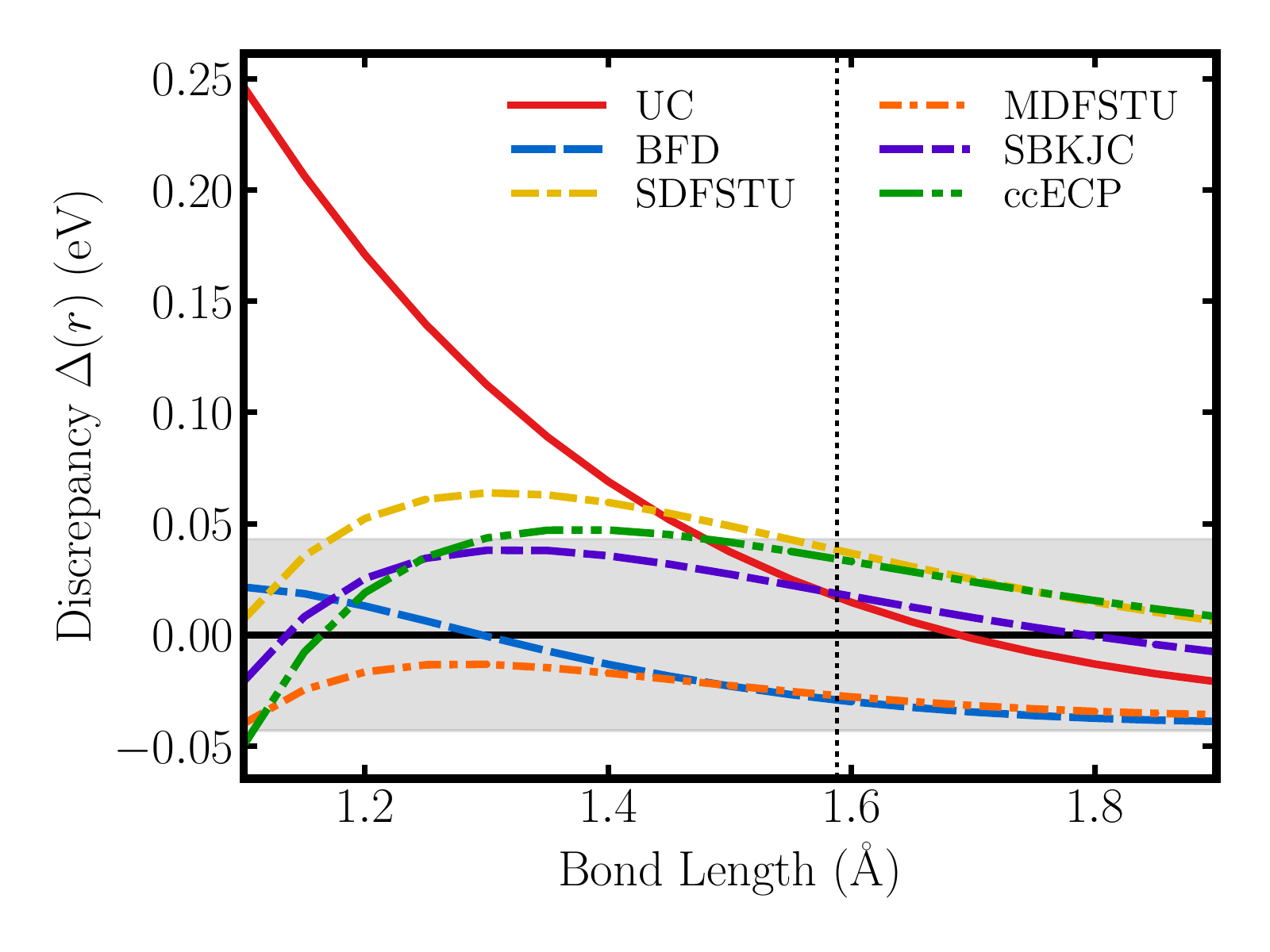}
\caption{GeH binding curve discrepancies}
\label{fig:GeH}
\end{subfigure}%
\begin{subfigure}{0.5\textwidth}
\includegraphics[width=\textwidth]{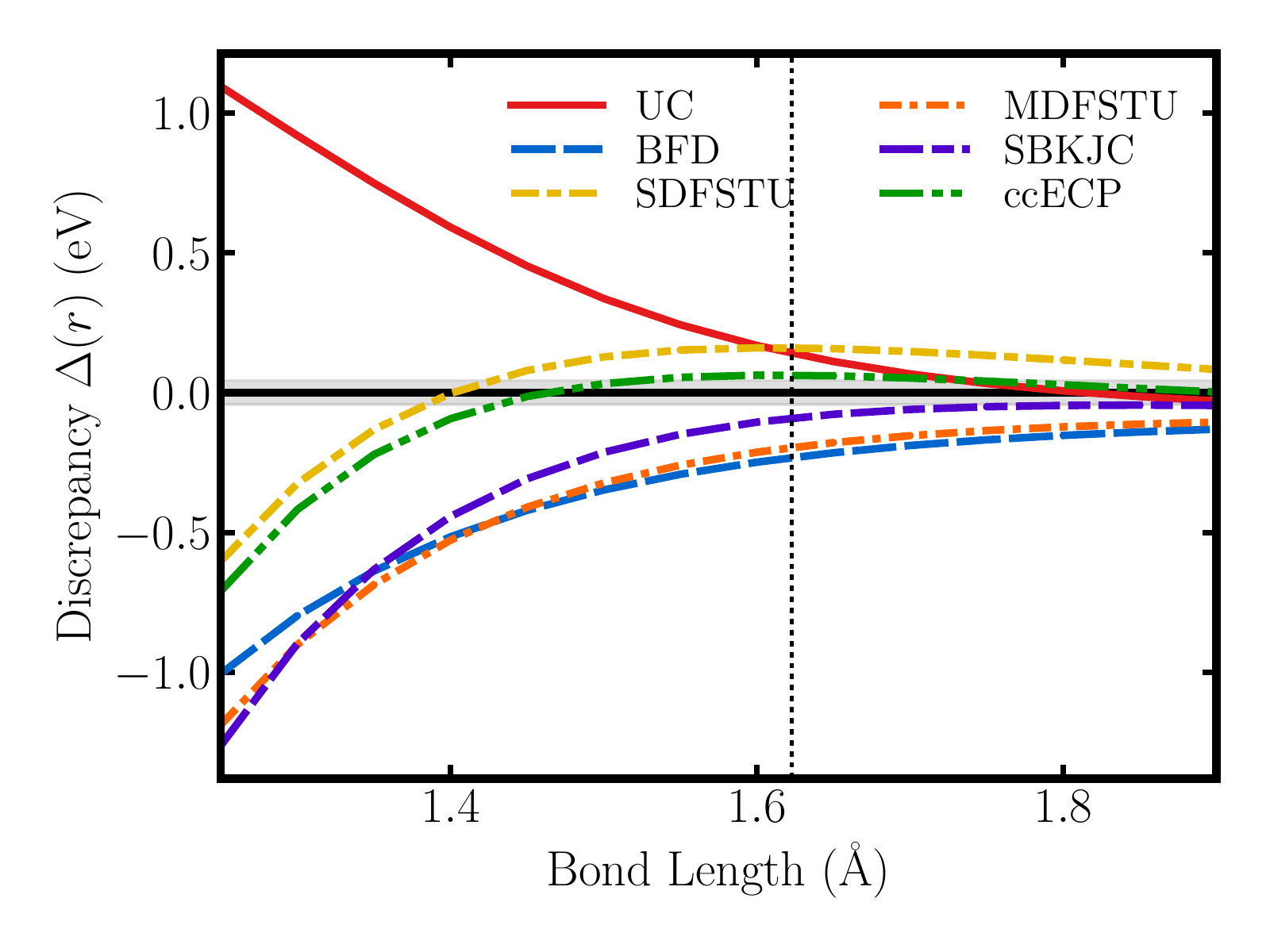}
\caption{GeO binding curve discrepancies}
\label{fig:GeO}
\end{subfigure}
\caption{Binding energy discrepancies for (a) GeH and (b) GeO molecules. 
The same notation applies as for the previous cases.}
\label{fig:Ge_mols}
\end{figure*}

\begin{figure*}[!htbp]
\centering
\begin{subfigure}{0.5\textwidth}
\includegraphics[width=\textwidth]{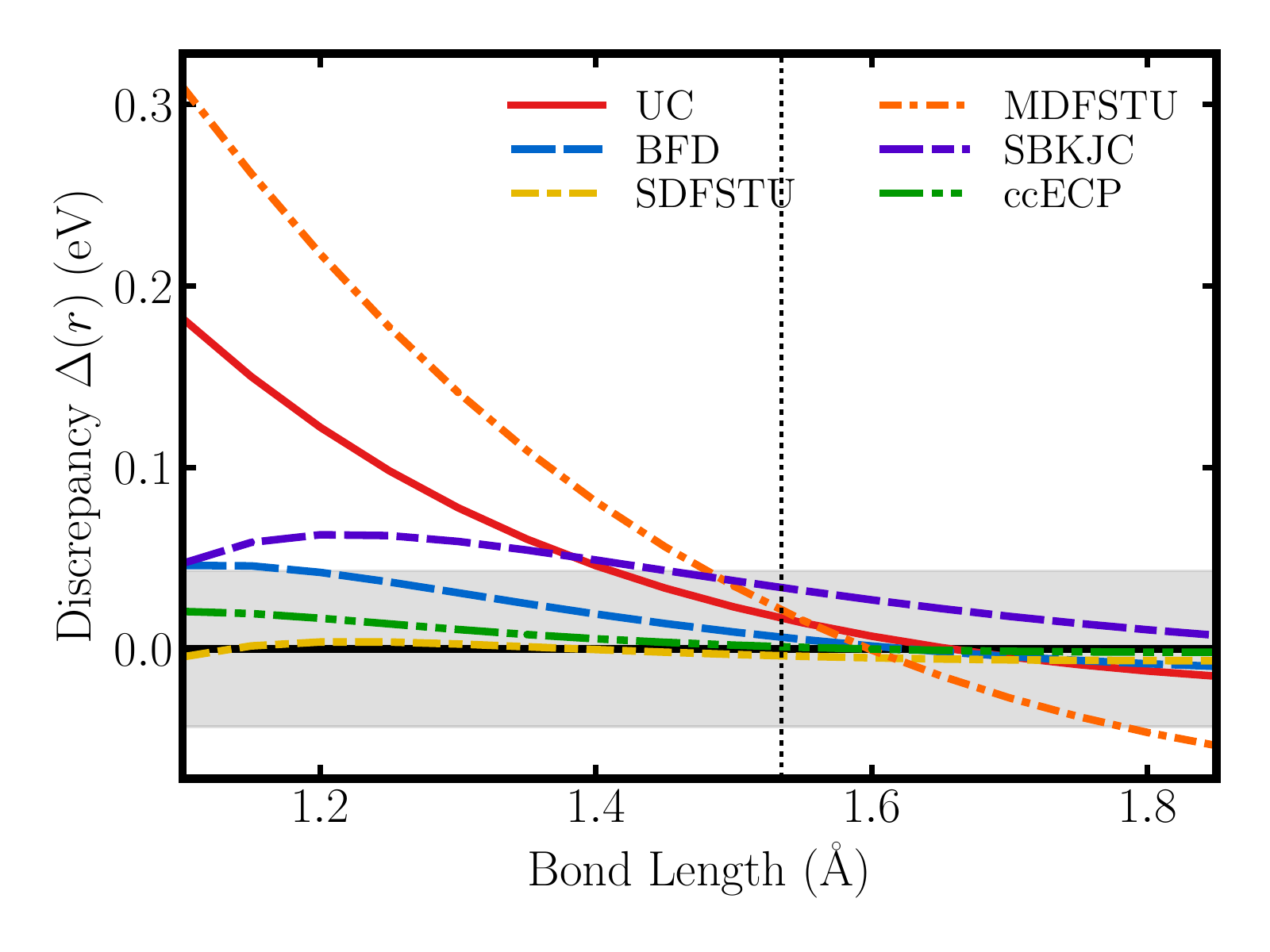}
\caption{AsH binding curve discrepancies}
\label{fig:AsH}
\end{subfigure}%
\begin{subfigure}{0.5\textwidth}
\includegraphics[width=\textwidth]{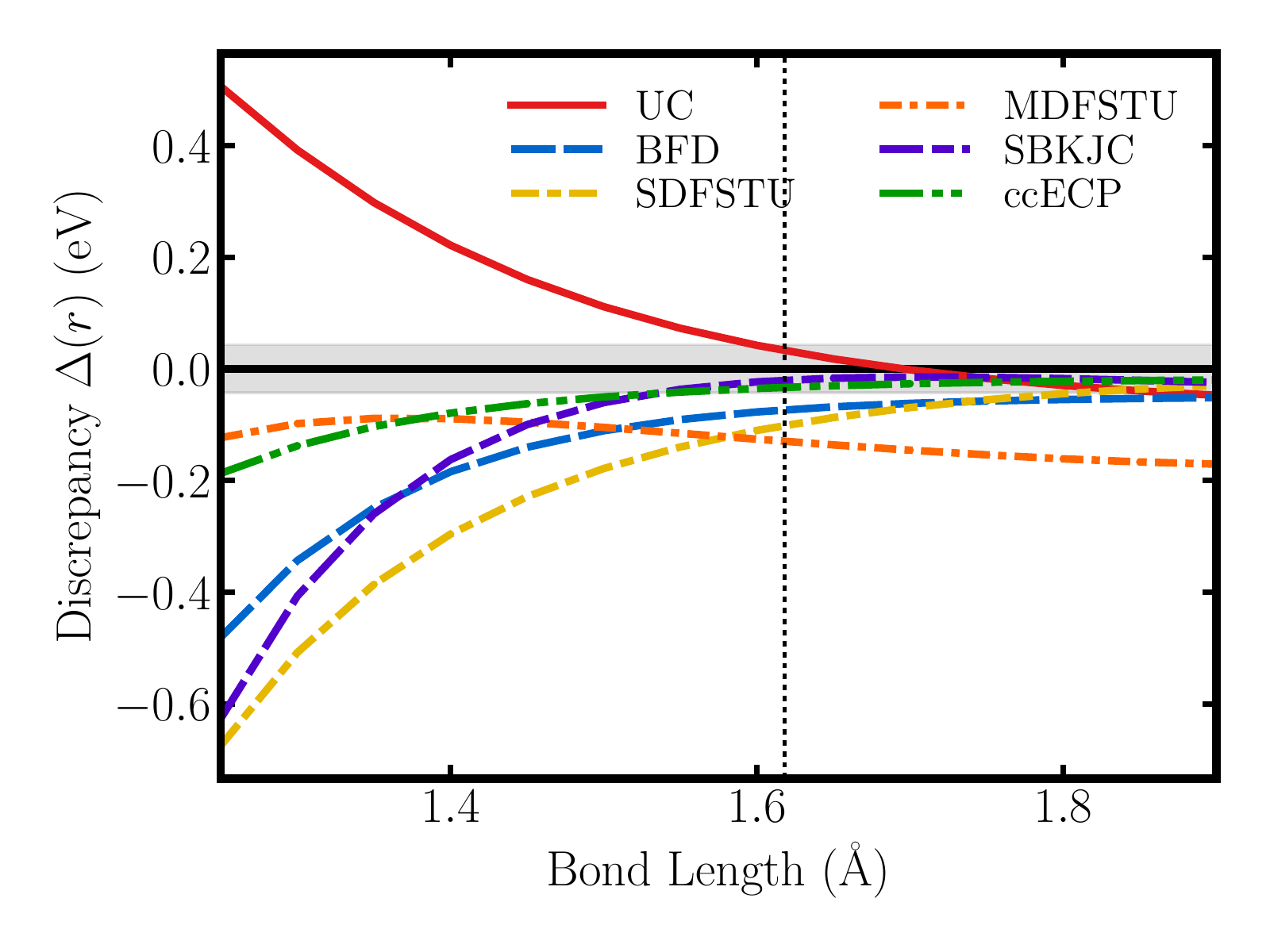}
\caption{AsO binding curve discrepancies}
\label{fig:AsO}
\end{subfigure}
\caption{Binding energy discrepancies for (a) AsH and (b) AsO molecules. The same notation applies as for the previous cases.}
\label{fig:As_mols}
\end{figure*}

\begin{figure*}[!htbp]
\centering
\begin{subfigure}{0.5\textwidth}
\includegraphics[width=\textwidth]{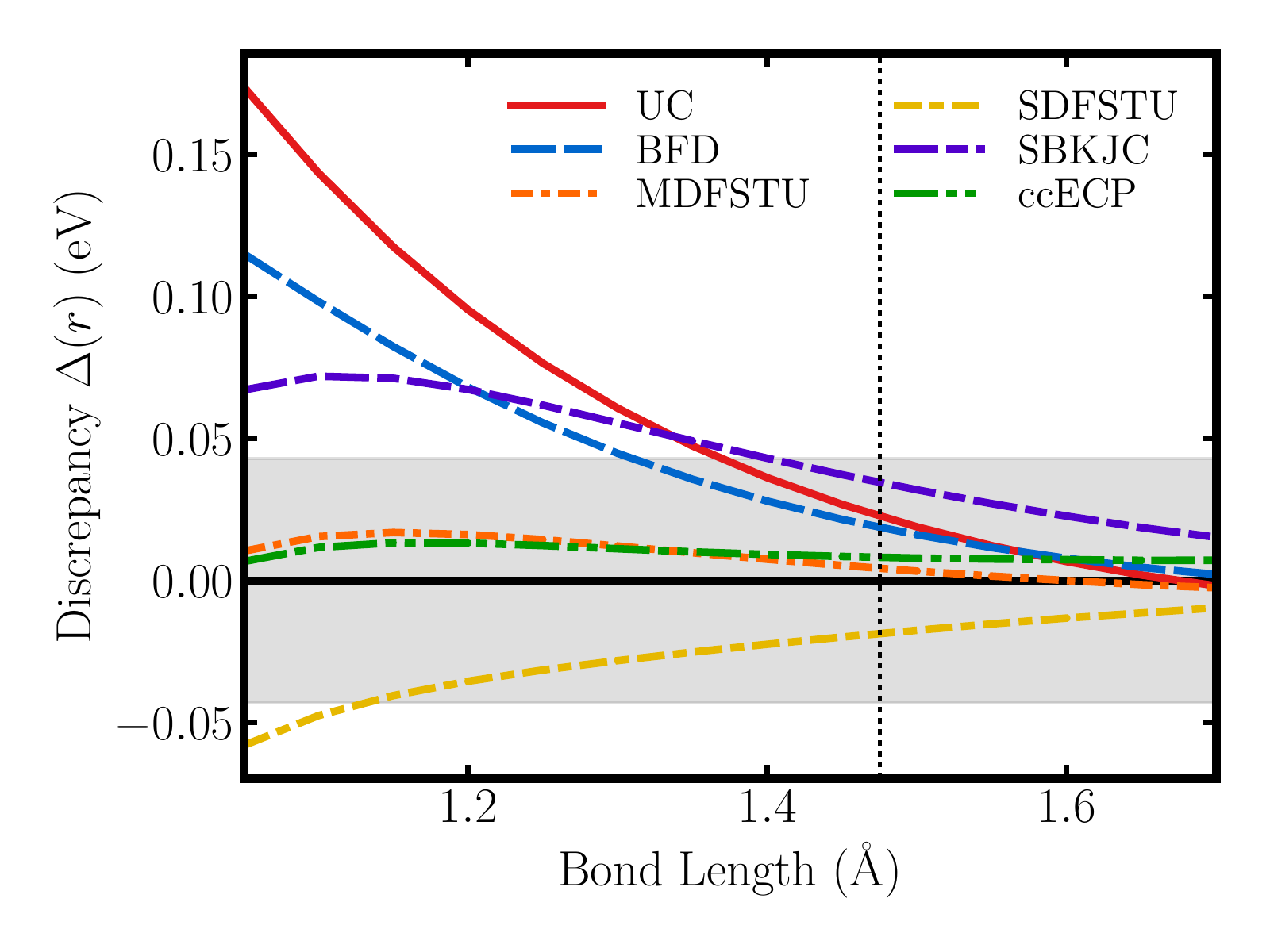}
\caption{SeH binding curve discrepancies}
\label{fig:SeH}
\end{subfigure}%
\begin{subfigure}{0.5\textwidth}
\includegraphics[width=\textwidth]{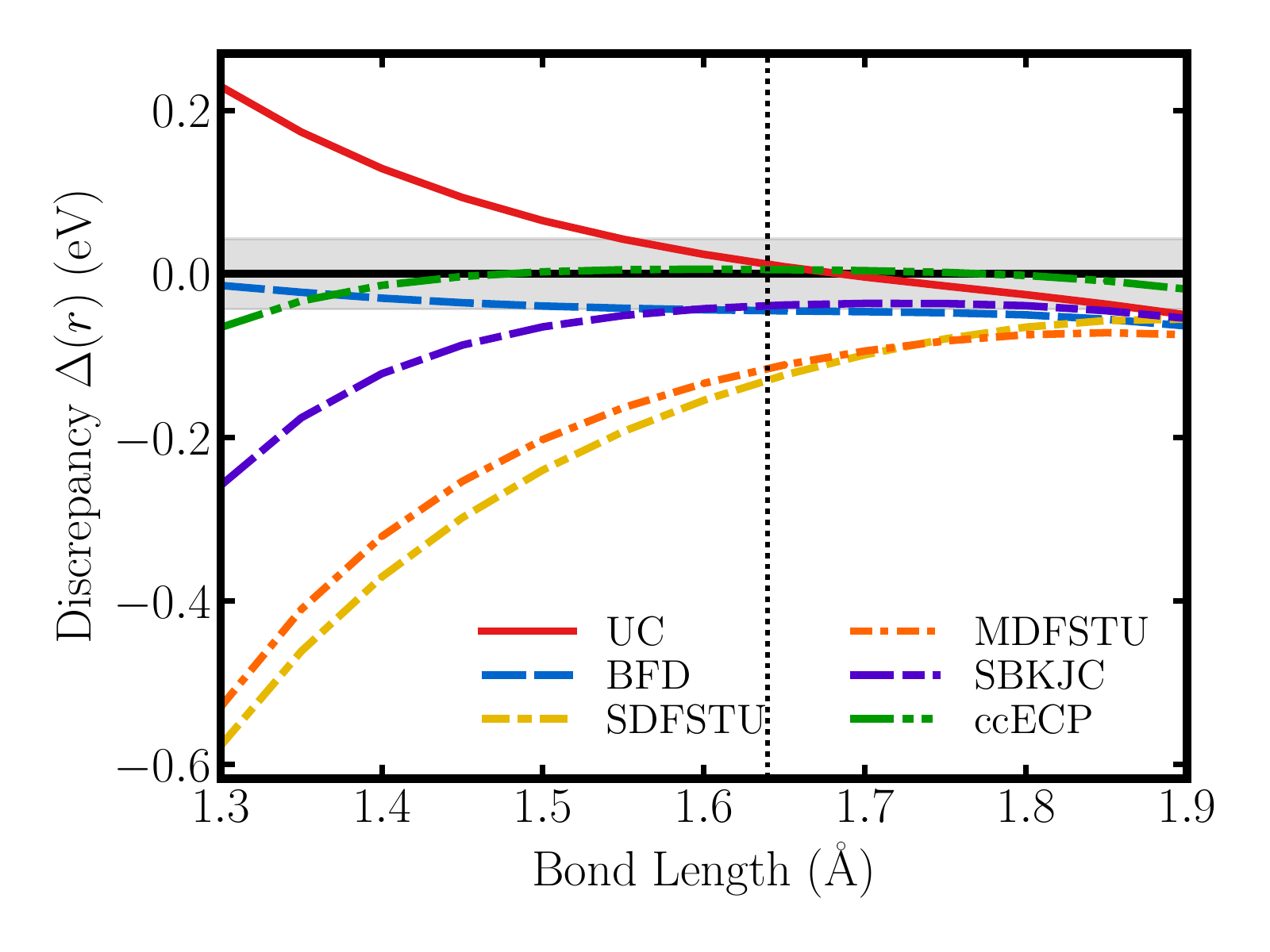}
\caption{SeO binding curve discrepancies}
\label{fig:SeO}
\end{subfigure}
\caption{Binding energy discrepancies for (a) SeH and (b) SeO molecules. 
The same notation applies as for the previous cases.
}
\label{fig:Se_mols}
\end{figure*}

\begin{figure*}[!htbp]
\centering
\begin{subfigure}{0.5\textwidth}
\includegraphics[width=\textwidth]{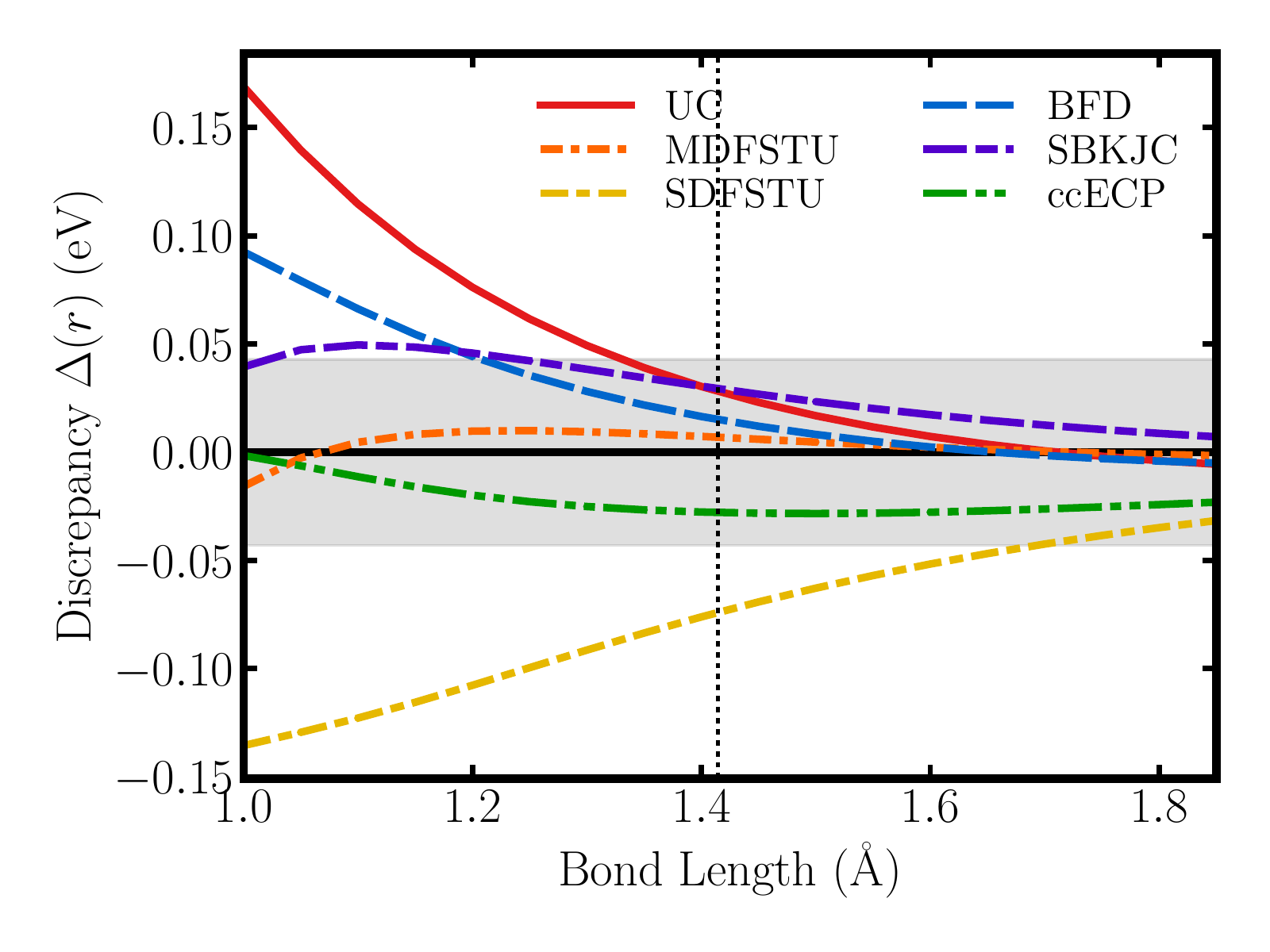}
\caption{BrH binding curve discrepancies}
\label{fig:BrH}
\end{subfigure}%
\begin{subfigure}{0.5\textwidth}
\includegraphics[width=\textwidth]{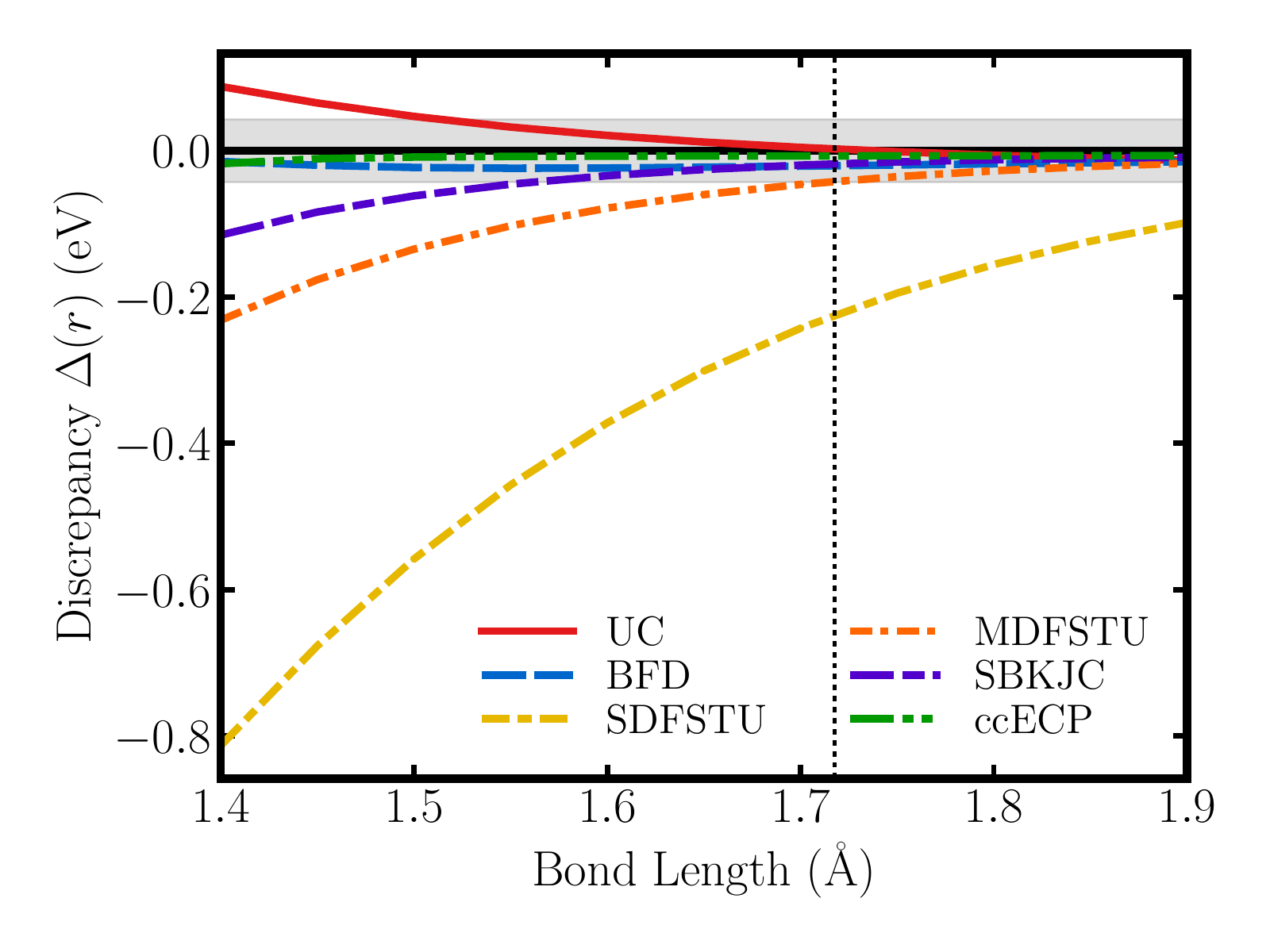}
\caption{BrO binding curve discrepancies}
\label{fig:BrO}
\end{subfigure}
\caption{Binding energy discrepancies for (a) BrH and (b) BrO molecules. 
The same notation applies as for the previous cases.
}
\label{fig:Br_mols}
\end{figure*}

\begin{figure}[!htbp]
\centering
\includegraphics[width=\columnwidth]{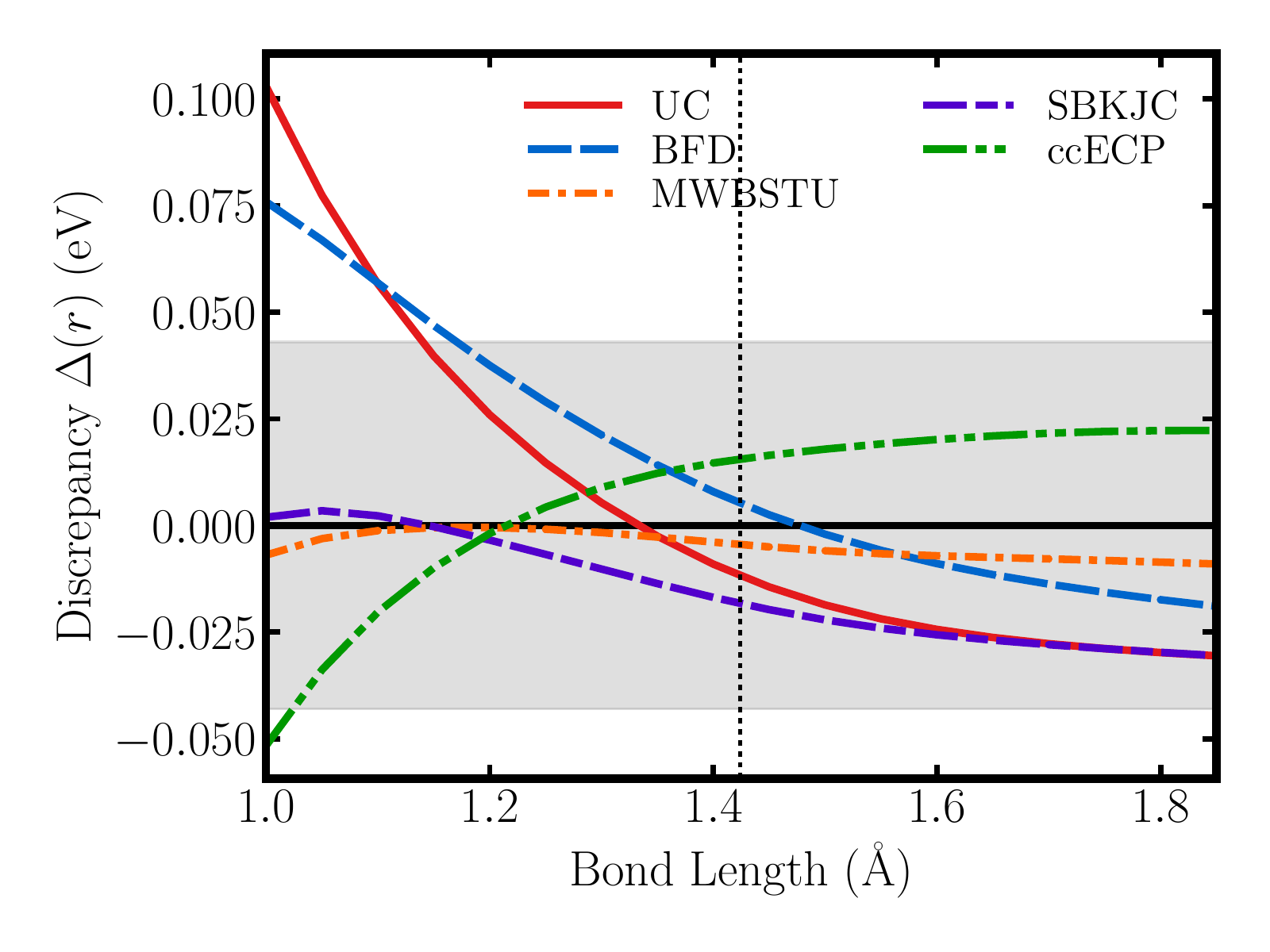}
\caption{Binding energy discrepancy for KrH$^+$ molecule. The binding curves are relative to the CCSD(T) binding curve. The shaded region indicates a discrepancy of chemical accuracy in either direction.}
\label{fig:KrH+}
\end{figure}

\begin{table}[!htbp]
\centering
\caption{Mean absolute deviations of binding parameters for various core approximations with respect to AE data for Ga-Kr hydride and oxide molecules. All parameters were obtained using Morse potential fit. The parameters shown are dissociation energy $D_e$, equilibrium bond length $r_e$, vibrational frequency $\omega_e$ and binding energy discrepancy at dissociation bond length $D_{diss}$.}
\label{morse:Ga-Kr}
\begin{tabular}{l|rrrrrrrrrr}
\hline\hline
{} & $D_e$(eV) & $r_e$(\AA) & $\omega_e$(cm$^{-1}$) & $D_{diss}$(eV) \\
\hline
UC         &  0.028(3) &  0.0092(7) &                 11(4) &        0.26(3) \\
BFD        &  0.093(4) &  0.0091(8) &                  9(4) &        0.34(3) \\
MDFSTU     &  0.066(3) &  0.0090(8) &                 25(3) &        0.27(3) \\
SBKJC      &  0.034(3) &  0.0052(6) &                 22(3) &        0.24(3) \\
SDFSTU     &  0.087(3) &  0.0097(8) &                 25(3) &        0.30(3) \\
ccECP      &  0.022(3) &  0.0030(8) &                 17(3) &        0.11(3) \\
\hline\hline
\end{tabular}
\end{table}

Table \ref{tab:MADs} shows the MADs and LMADs of the considered spectrum for Ga - Kr atoms. 
For each element, we choose a consistent set of excitations/gaps relative to the ground state which represents the spectrum. 
The spectrum consists of electron affinity, all ionization potentials down to single electron in the valence space, $4s$ $\rightarrow$ $4p$ excited states and various $4d$ occupied states. 

In all cases, our ccECP has lower MAD \textit{and} LMAD compared to all core approximations including uncorrelated-core (UC).
The only exception to this is Ge where SBKJC achieves lower MAD although it is higher in LMAD while SDFSTU achieves lower LMAD but is higher in MAD.
We believe the ccECP is perhaps the most reasonable compromise between these two; this becomes clearer by looking at molecular discrepancies (figure \ref{fig:Ge_mols}).
Speaking in general terms, significant gains in accuracy are obtained for As, Se, Br, and Kr elements while Ga and Ge show modest improvements or are on par with previously published constructions. 
However, we are pleased to see more consistent improvements in molecular properties in our ccECPs as follows.    

We show the transferability results for these elements in figures \ref{fig:Ga_mols}, \ref{fig:Ge_mols}, \ref{fig:As_mols}, \ref{fig:Se_mols}, \ref{fig:Br_mols}, \ref{fig:KrH+} which are hydride and oxide molecules for Ga, Ge, As, Se, Br, Kr atoms respectively.
For GaO molecule, we omit the UC molecular binding discrepancy curve where we were unable to converge the coupled cluster molecular calculations correctly. 
For Kr, we calculate the KrH$^+$ discrepancies only.
Similarly to the picture seen in the MAD of the spectrum, the ccECP has the lowest errors in molecular properties when compared to all other core approximations.
We observe improvements in Ga and Ge molecules while As molecular properties show more notable gains in accuracy. 
Significant improvements are seen in Se, Br, and Kr where the ccECP is essentially always within chemical accuracy for both hydrides and oxides in all bond lengths up to dissociation limit.
The summary of MADs of Morse potential fit parameters can be seen in table \ref{morse:Ga-Kr}.
We can see that our ccECP has the smallest errors on $D_e$, $r_e$ and $D_{diss}$ when compared to other ECPs.
Only BFD has a smaller error in $\omega_e$, however, this result is not fully consistent with errors for the other parameters being notably larger.

In these $4p$ main group elements, we observe that the achieved ccECP MADs are significantly higher when compared to K and Ca.  Also, we were not able to achieve chemical accuracy in some compressed oxides although all ccECP hydrides are essentially within 1 kcal/mol.
This is mainly due to the smaller size of the valence space.
If higher accuracy is desired, a smaller core with a larger valence space will be required, but it is beyond the scope of this work. 

\subsection{Additional pseudopotentials: H, He, Li, Be, F, and Ne}\label{sec:additional}

\begin{table}[htbp!]
\setlength{\tabcolsep}{4pt} 
\small
\centering
\caption{ Parameter values for additional ccECPs. For all ECPs, the highest $\ell$ value corresponds to the local channel $\ell_{loc}$. 
Note that the highest non-local angular momentum channel $\ell_{max}$ is related to it as $\ell_{max}=\ell_{loc}-1$.
ECPs indicated with "reg" do not remove any electrons and contain only local channel.}
\label{tab:additional_params}
\begin{tabular}{ccrrrrrccrrrrr}
\hline\hline
\multicolumn{1}{c}{Atom} & \multicolumn{1}{c}{$Z_{\rm eff}$} & \multicolumn{1}{c}{$\ell$} & \multicolumn{1}{c}{$n_{\ell k}$} & \multicolumn{1}{c}{$\alpha_{\ell k}$} & \multicolumn{1}{c}{$\beta_{\ell k}$}  \\
\hline

H        & 1 & 0 &  2 &  1.000000 &   0.000000    \\
         &   & 1 &  1 & 21.243595 &   1.000000    \\
         &   & 1 &  3 & 21.243595 &  21.243595    \\
         &   & 1 &  2 & 21.776967 & -10.851924    \\
         &&&&&                                    \\
He       & 2 & 0 &  2 &  1.000000 &   0.000000    \\
         &   & 1 &  1 & 32.000000 &   2.000000    \\
         &   & 1 &  3 & 32.000000 &  64.000000    \\
         &   & 1 &  2 & 33.713355 & -27.700840    \\
         &&&&&                                    \\
Li(reg)  & 3 & 0 &  2 &  10.00000 &   0.000000    \\
         &   & 1 &  1 &  24.00000 &   3.000000    \\
         &   & 1 &  3 &  24.00000 &  72.000000    \\
         &   & 1 &  2 &  24.00000 & -36.443940    \\
         &   & 1 &  2 &  9.136790 &   0.780200    \\
         &&&&&                                    \\
Be(reg)  & 4 & 0 &  2 & 10.000000 &   0.000000    \\
         &   & 1 &  1 & 18.441179 &   4.000000    \\
         &   & 1 &  3 & 22.941215 &  73.764716    \\
         &   & 1 &  2 & 30.142700 &  -52.33020    \\
         &   & 1 &  2 & 11.327931 &  -3.360040    \\
         &&&&&& \\
Li &  1 & 0 &  2 &  1.330248 &  6.752868          \\
   &    & 1 &  1 & 15.000000 &  1.000000          \\
   &    & 1 &  3 & 15.047997 & 15.000000          \\
   &    & 1 &  2 &  1.806054 & -1.242730          \\
   &&&&&                                          \\
Be &  2 & 0 &  2 &  2.487404 &  12.663919         \\
   &    & 1 &  1 & 17.949002 &   2.000000         \\
   &    & 1 &  3 & 24.132003 &  35.898004         \\
   &    & 1 &  2 & 20.138003 & -12.774998         \\
   &    & 1 &  2 &  4.333171 &  -2.960014         \\
   &&&&&   \\                                    
F  &  7 & 0 &  2 & 14.780765 &  78.901772         \\ 
   &    & 1 &  1 & 12.087585 &   7.000000         \\ 
   &    & 1 &  3 & 12.838063 &  84.613094         \\ 
   &    & 1 &  2 & 12.312346 & -53.027517         \\ 
   &&&&&                                          \\ 
Ne &  8 &  0 &  2 & 16.554415 &  81.622057        \\ 
   &    &  1 &  1 & 14.793512 &   8.000000        \\ 
   &    &  1 &  3 & 16.582039 & 118.348096        \\ 
   &    &  1 &  2 & 16.080735 & -70.278859        \\ 
   &&&&&                                          \\ 
\hline\hline
\end{tabular}
\end{table}

\begin{table*}
\centering
\caption{A summary of the spectrum discrepancies of various ECPs for selected elements. For each atom, we provide MAD of the entire calculated spectrum and MAD of selected low-lying atomic states (LMAD). The LMAD atomic states include electron affinity, first and second ionizations only. All values are in eV.}
\label{tab:MAD_additional}
\begin{tabular}{lcccccccccccccc}
\hline\hline
Atom &  Quantity   &   UC &    BFD &    SDFSTU   &    MWBSTU   &    eCEPP &     SBKJC &  CRENBL   &  ccECP  \\
\hline
Li         & MAD &  0.026002 &  0.028279 &  0.033749  &            & 0.034373 &  0.027706 & 0.025479 &  0.024961 \\
\hline
Be         & MAD &  0.039119 &  0.039855 &  0.039200  &            & 0.041993 &  0.038804 & 0.043981 &  0.032178 \\
\hline   
F          & MAD &  0.087945 &  0.266950 &  0.187671  &  0.217170  & 0.021450 &  0.015619 & 0.018112 &  0.028159 \\
           &LMAD &  0.028785 &  0.080417 &  0.025233  &  0.033469  & 0.007550 &  0.005818 & 0.005222 &  0.004284 \\
\hline    
Ne         & MAD &  0.124251 &  0.320135 &            &  0.212618  &          &  0.043077 & 0.036869 &  0.023995 \\
           &LMAD &  0.027683 &  0.055655 &            &  0.027137  &          &  0.014303 & 0.006541 &  0.002008 \\
\hline
\hline
\end{tabular}
\end{table*}

In this section, we provide the data for several lighter elements that were not included  in our previously published sets to
complete our ccECP table for elements H through Kr. 
The elements considered here are H, He, Li, Be, F and Ne.


\begin{figure*}[!htbp]
\centering
\begin{subfigure}{0.5\textwidth}
\includegraphics[width=\textwidth]{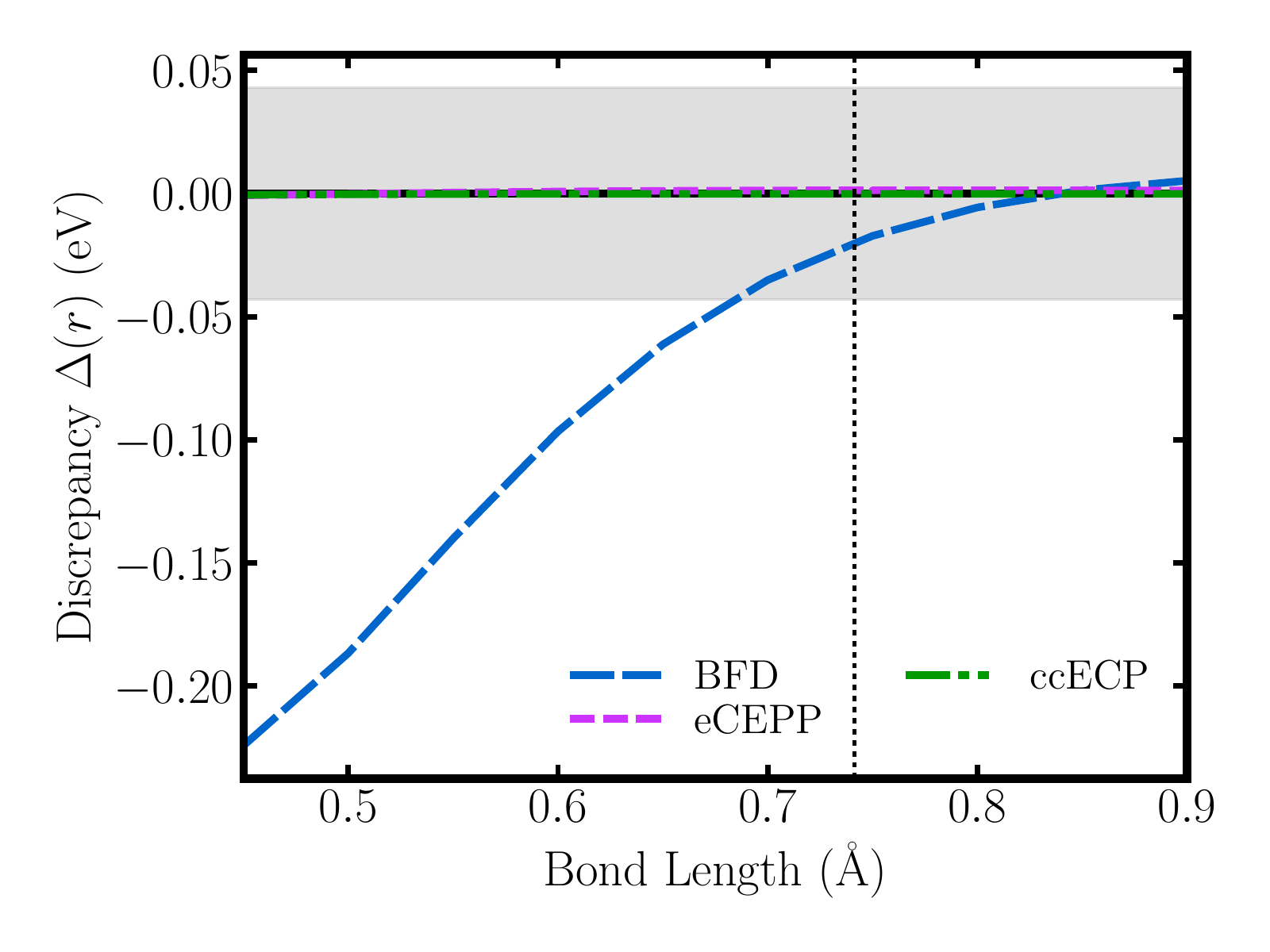}
\caption{Binding energy discrepancies for H$_2$ molecule}
\label{fig:H2}
\end{subfigure}%
\begin{subfigure}{0.5\textwidth}
\includegraphics[width=\textwidth]{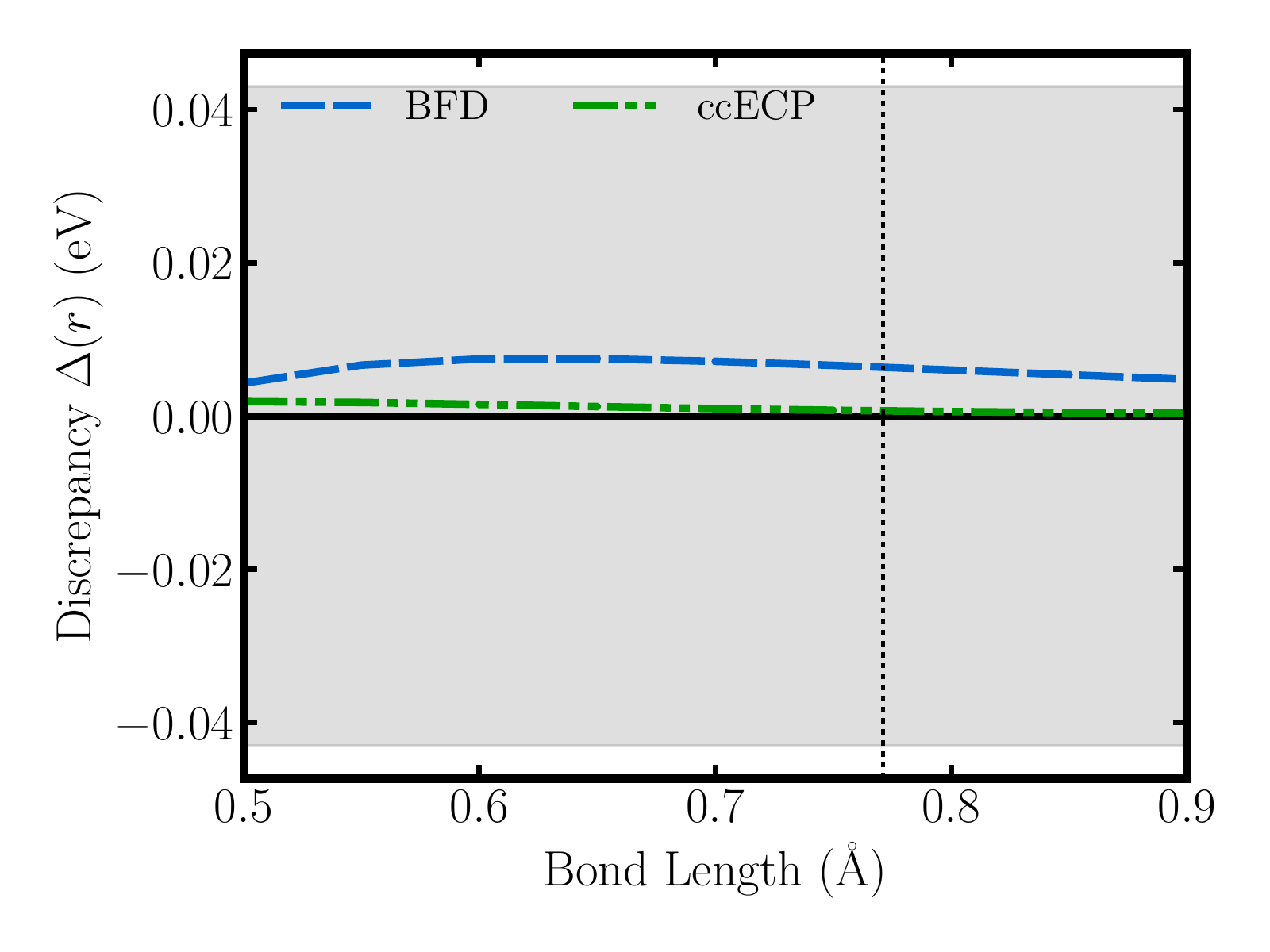}
\caption{Binding energy discrepancies for HeH$^+$ molecule}
\label{fig:HeH}
\end{subfigure}
\caption{Binding energy discrepancies for (a)  H$_2$ and (b) HeH$^+$ molecules. 
The same notation applies as for the previous cases.
}
\label{fig:nocore_mols}
\end{figure*}

For H, He, Li and Be, we developed local pseudopotentials, where there are no non-local operators and no electrons are removed.
These all-electron ECPs are used to cancel the Coulomb singularity which can be useful for computational efficiency purposes, thus resulting in soft, regularized potentials.
We indicate these ECPs by label "reg" whenever there is ambiguity. 

Although H ccECP was published before \cite{3-ccECP}, we provide the missing molecular tests here.
We show the quality of regularized H and He ECPs by transferability tests on H$_2$ and HeH$^+$ molecules which are shown in figure \ref{fig:nocore_mols}.
Note that the originally published BFD ECP significantly over-binds the H$_2$ molecule due to a rather significant radial span
of the regularized Coulomb potential.


\begin{figure*}[!htbp]
\centering
\begin{subfigure}{0.5\textwidth}
\includegraphics[width=\textwidth]{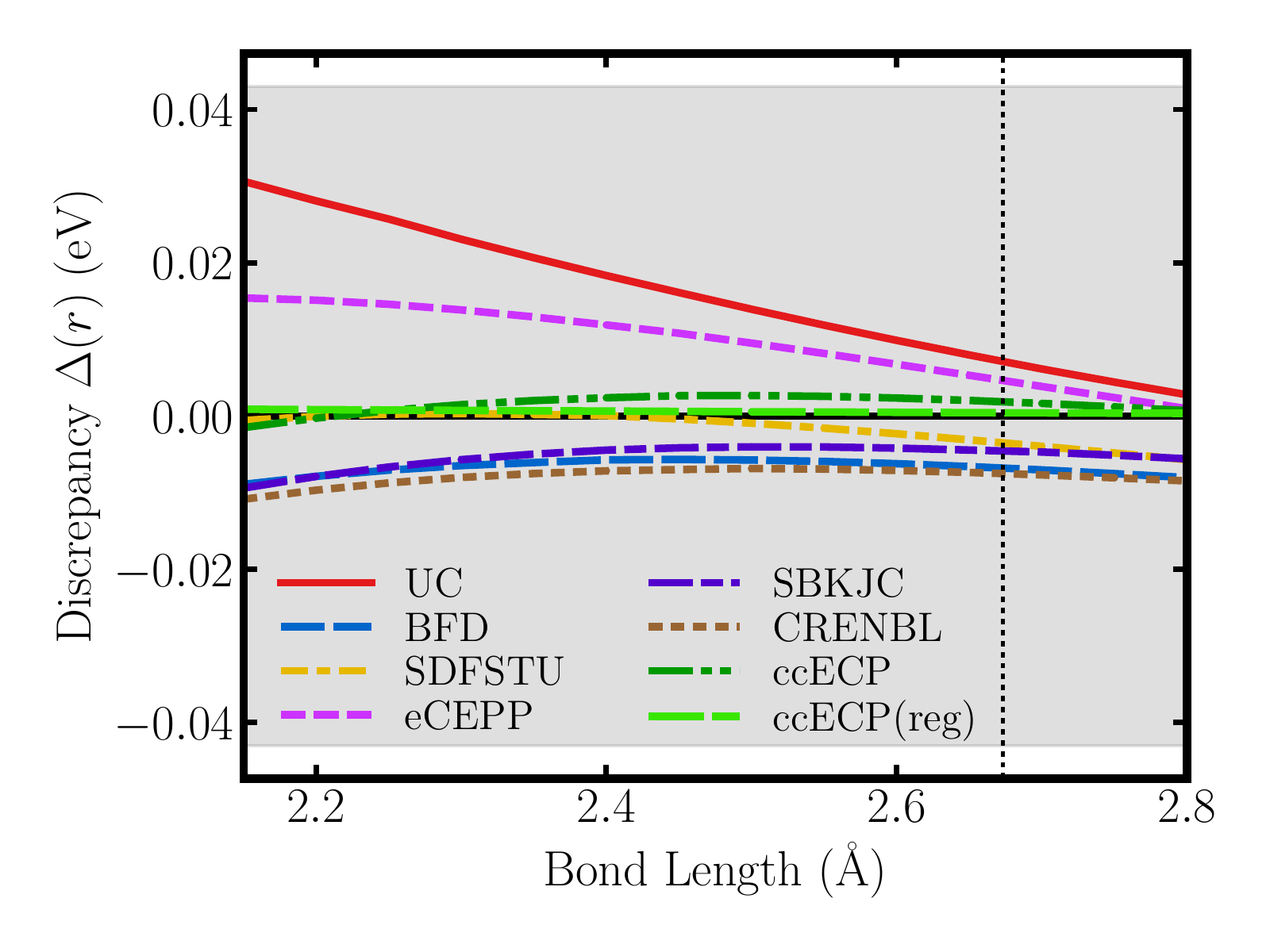}
\caption{Li$_2$ binding curve discrepancies}
\label{fig:Li2}
\end{subfigure}%
\begin{subfigure}{0.5\textwidth}
\includegraphics[width=\textwidth]{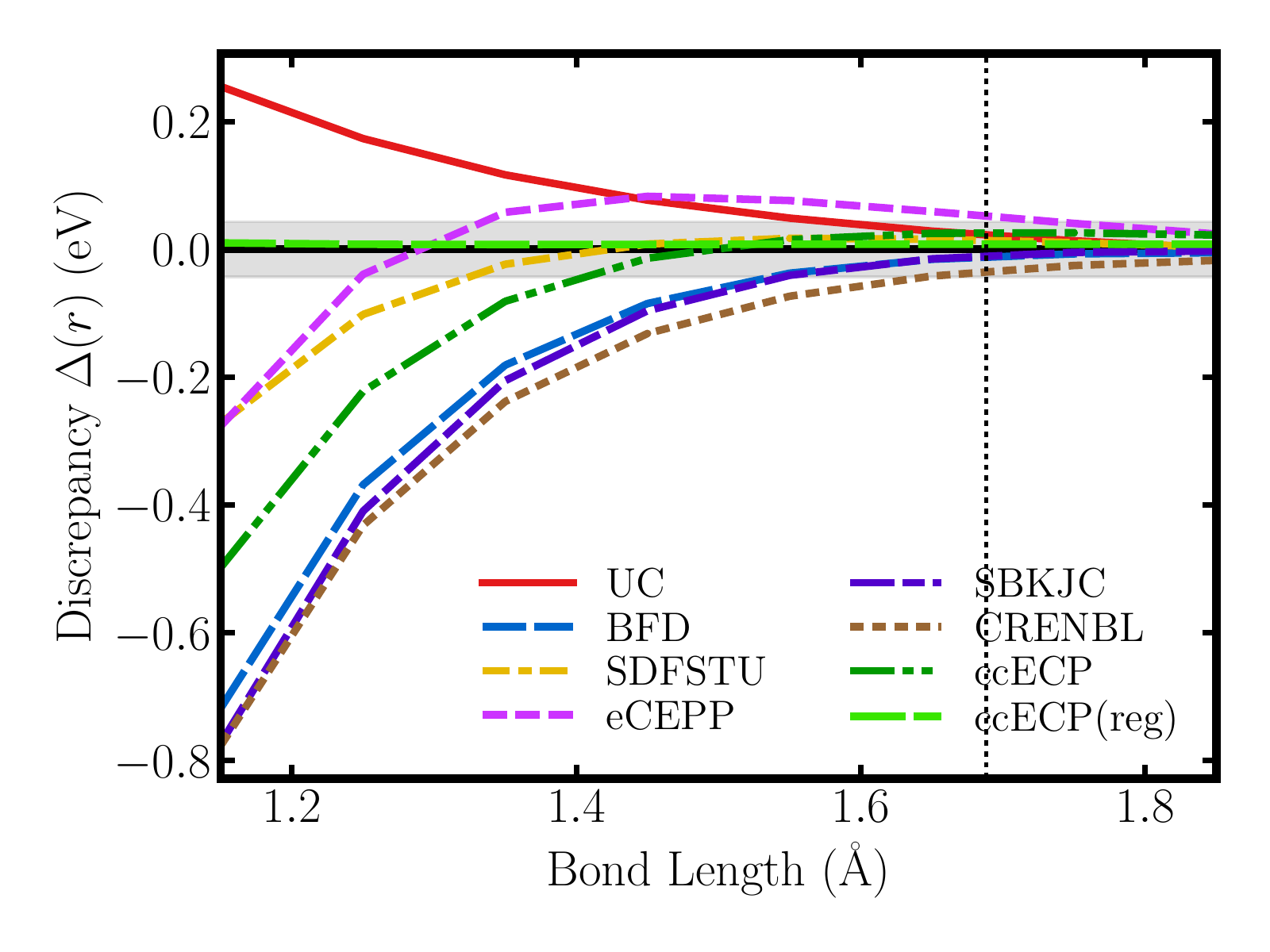}
\caption{LiO binding curve discrepancies}
\label{fig:LiO}
\end{subfigure}
\caption{Binding energy discrepancies for (a) Li$_2$ and (b) LiO molecules. 
QZ basis and ccECP oxygen was used for all cases.
Otherwise the same notation applies as for the previous cases.
}
\label{fig:Li_mols}
\end{figure*}

\begin{figure*}[!htbp]
\centering
\begin{subfigure}{0.5\textwidth}
\includegraphics[width=\textwidth]{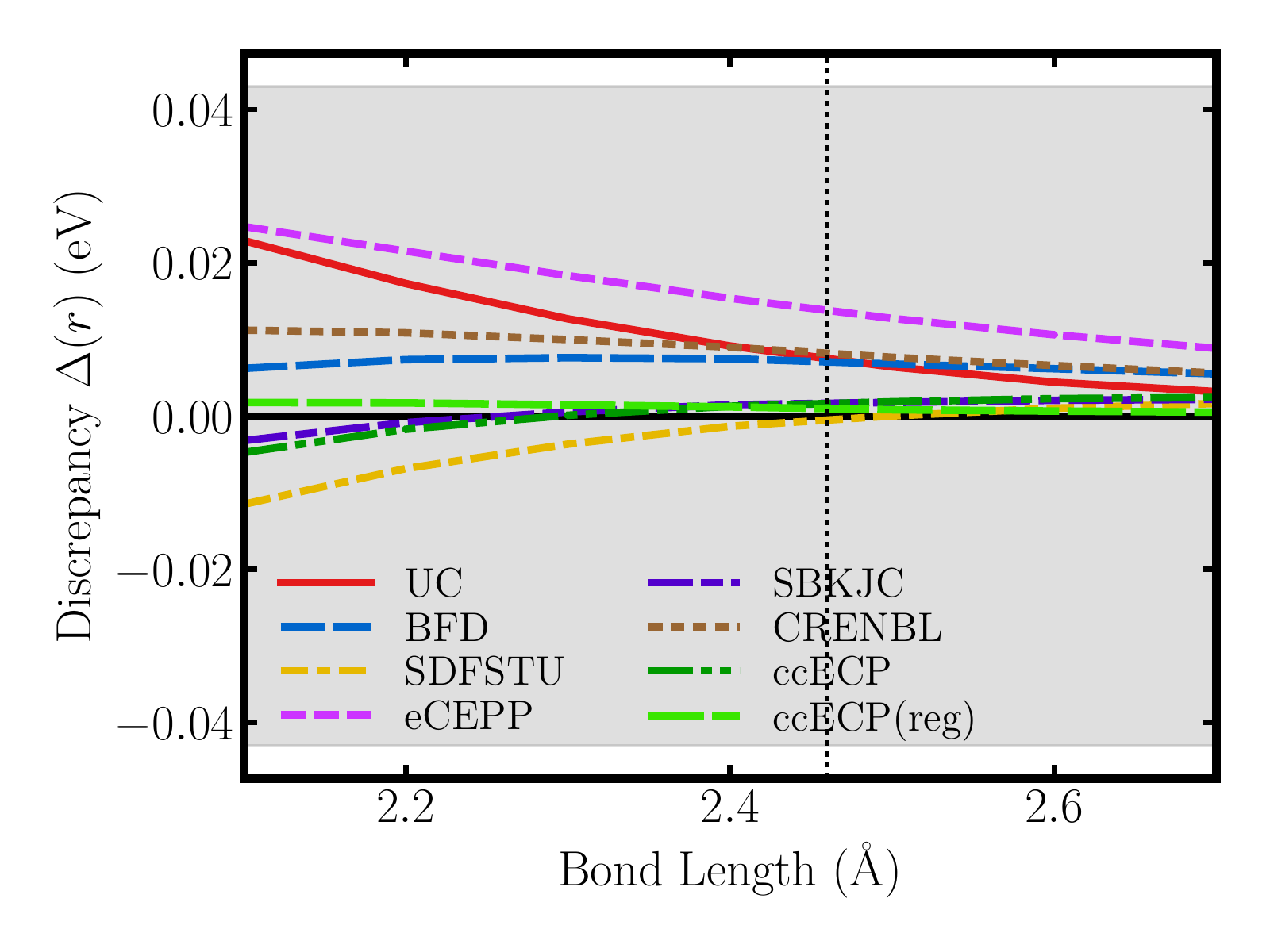}
\caption{Be$_2$ binding curve discrepancies}
\label{fig:Be2}
\end{subfigure}%
\begin{subfigure}{0.5\textwidth}
\includegraphics[width=\textwidth]{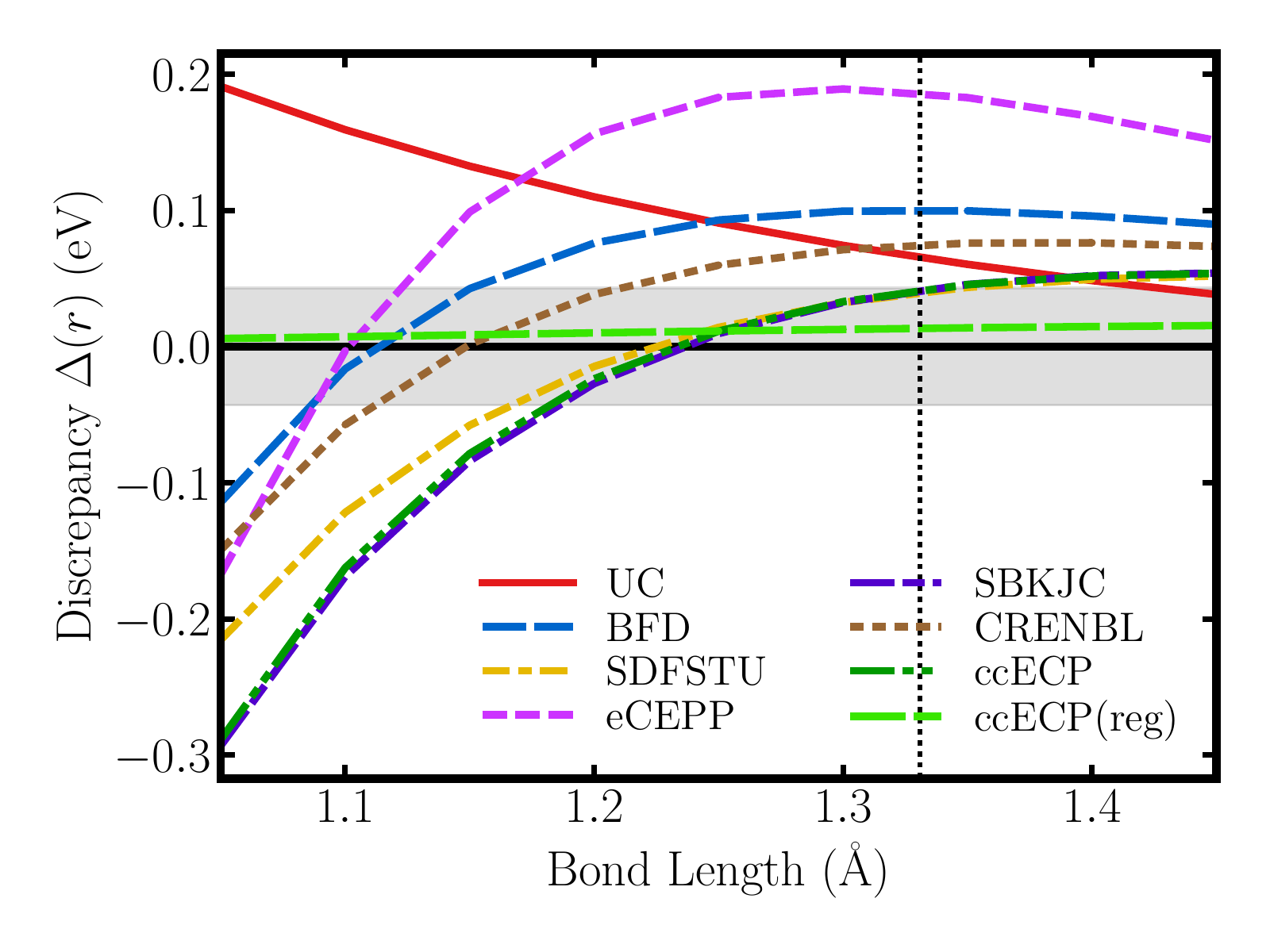}
\caption{BeO binding curve discrepancies}
\label{fig:BeO}
\end{subfigure}
\caption{Binding energy discrepancies for (a) Be$_2$ and (b) BeO molecules. QZ basis and ccECP oxygen was used for all cases.
Otherwise the same notation applies as for the previous cases.}
\label{fig:Be_mols}
\end{figure*}

For Li and Be, we also developed He-core ECPs with an $s$ non-local channel apart from all-electron pseudopotentials.
The spectrum and molecular tests for Li and Be are given in table \ref{tab:MAD_additional} and figures \ref{fig:Li_mols}, \ref{fig:Be_mols} respectively.
The ccECPs have the most accurate spectrum for both elements among all core approximations.
ccECP(reg) improves it significantly further by achieving MAD=0.00042 eV for Li and MAD=0.00195 eV for Be.

Note that both Li and Be dimer curves are perfectly reproduced by ccECPs although oxides show considerable errors, as for all other ECPs.
SDFSTU however, is distinctly better near dissociation region in LiO and BeO.
We estimate that this is due to the ECP parameterization and the constraints of smoothness of our ECP form.
If higher accuracy is desired, our ccECP(reg) is essentially flat and is always within 0.015 eV error in Li and Be dimers and oxides.


\begin{figure*}[!htbp]
\centering
\begin{subfigure}{0.5\textwidth}
\includegraphics[width=\textwidth]{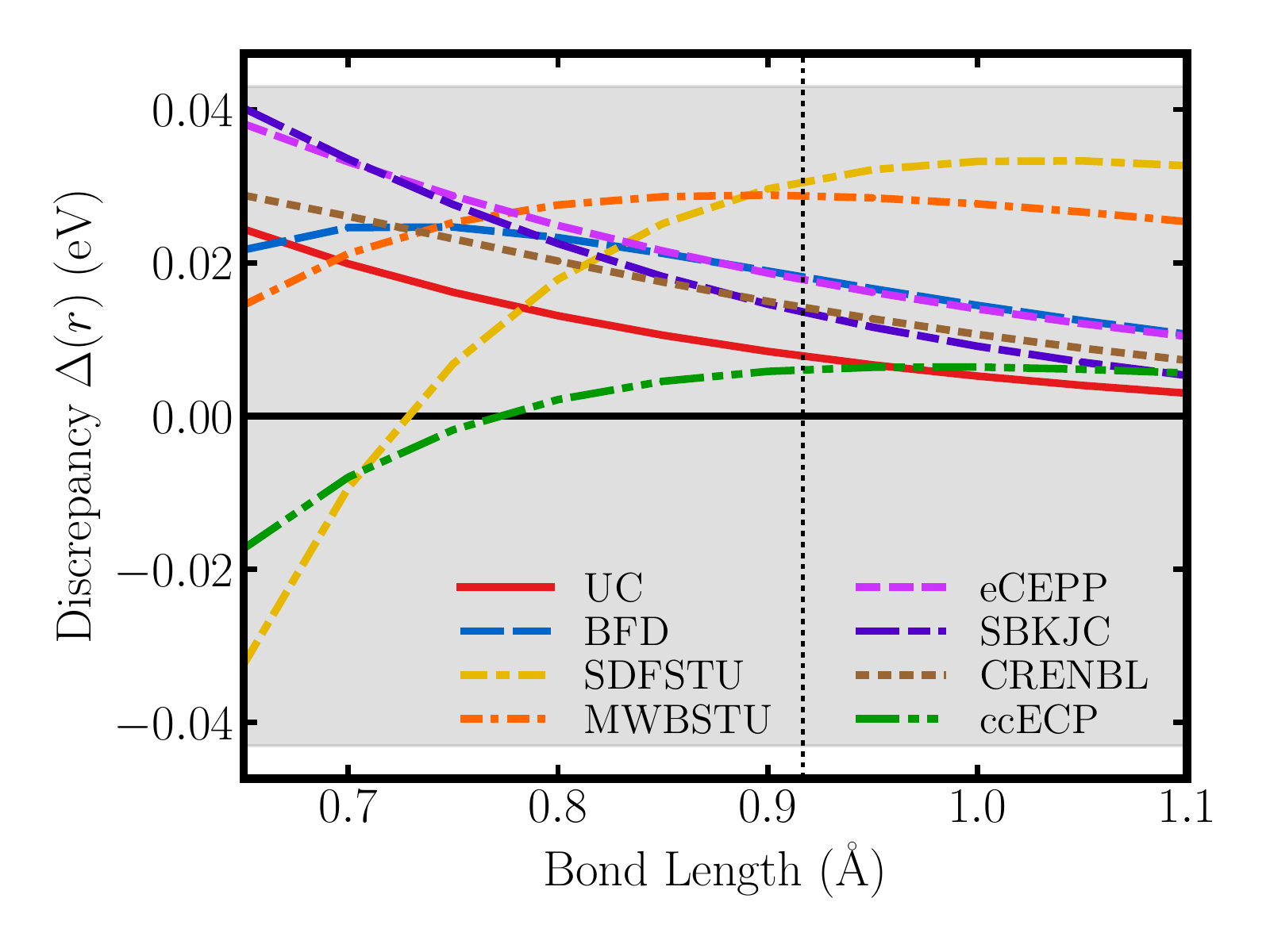}
\caption{FH binding curve discrepancies}
\label{fig:FH}
\end{subfigure}%
\begin{subfigure}{0.5\textwidth}
\includegraphics[width=\textwidth]{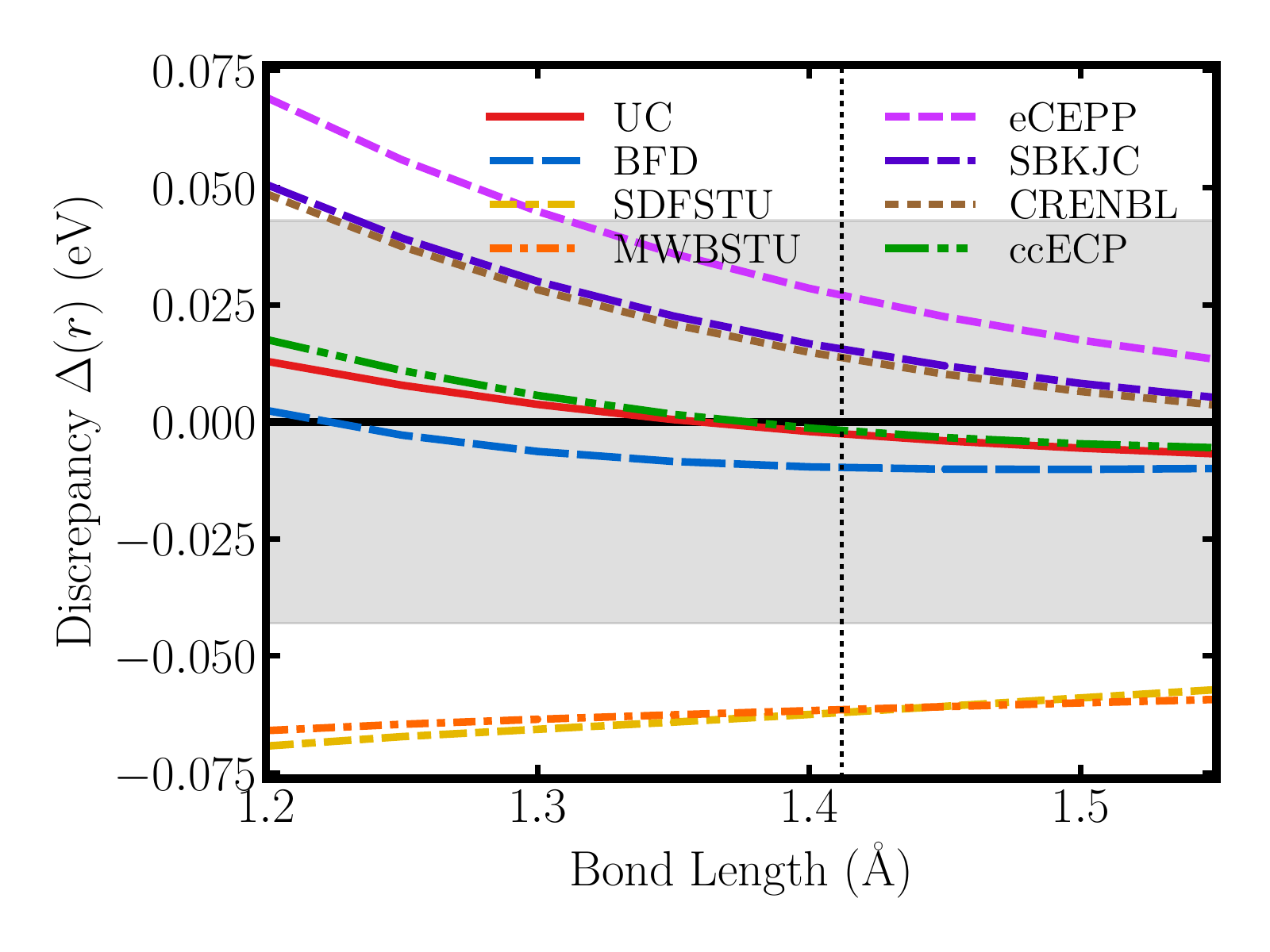}
\caption{F$_2$ binding curve discrepancies}
\label{fig:F2}
\end{subfigure}
\caption{Binding energy discrepancies for (a) FH and (b) F$_2$ molecules. 
The same notation applies as for the previous cases. }
\label{fig:F_mols}
\end{figure*}

\begin{figure}[htbp!]
\caption{Binding energy discrepancies for NeH$^+$ molecule. The same notation applies as for the previous cases.}
\includegraphics[width=\columnwidth]{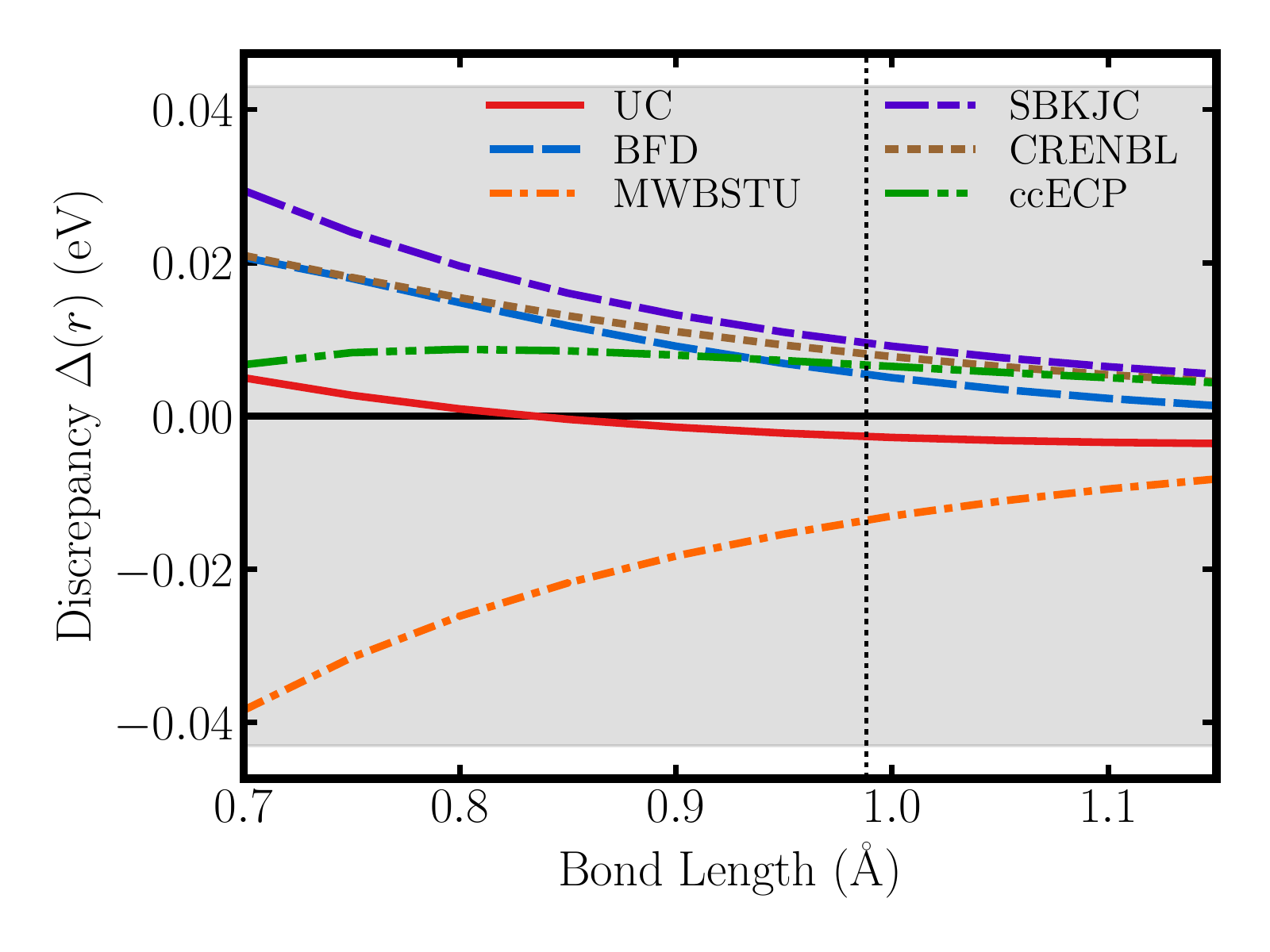}
\label{fig:NeH}
\centering
\end{figure}

For F and Ne, we use He-core ECPs with an $s$ non-local channel.
The spectrum MADs for F and Ne atoms are shown in table \ref{tab:MAD_additional}.
In F, ccECP has the lowest LMAD, although eCEPP, SBKJC and CRENBL have lower MAD.
However, they notably underbind in F$_2$ molecule whereas ccECP is close AE values as can be seen in figure \ref{fig:F_mols}.
In Ne, the ccECP achieves the lowest MAD \textit{and} LMAD in the spectrum while achieving a more constant and overall smaller error in NeH$^+$ molecule, shown in figure \ref{fig:NeH}.


\begin{table}[!htbp]
\centering
\caption{Mean absolute deviations of binding parameters for various core approximations with respect to AE data for  H, He, Li, Be, F, and Ne related molecules. All parameters were obtained using Morse potential fit. The parameters shown are dissociation energy $D_e$, equilibrium bond length $r_e$, vibrational frequency $\omega_e$ and binding energy discrepancy at dissociation bond length $D_{diss}$.}
\label{morse:additional}
\begin{tabular}{l|rrrrrrrrrr}
\hline\hline
{} & $D_e$(eV) & $r_e$(\AA) & $\omega_e$(cm$^{-1}$) & $D_{diss}$(eV) \\
\hline
UC             &  0.018(4) &   0.011(2) &     6.2(8.8)    &        0.09(5) \\
BFD            &  0.023(3) &   0.005(1) &       21(12)    &        0.15(5) \\
CRENBL         &  0.027(3) &   0.007(2) &    15.7(8.8)    &        0.18(4) \\
SBKJC          &  0.016(3) &   0.006(2) &    15.5(8.8)    &        0.20(4) \\
SDFSTU         &  0.027(4) &   0.006(2) &    14.9(9.1)    &        0.12(5) \\
eCEPP          &  0.047(4) &   0.011(2) &       30(13)    &        0.11(6) \\
ccECP          &  0.010(3) &   0.003(1) &       12(11)    &        0.11(5) \\
ccECP(reg)     &  0.006(4) &   0.001(3) &     0.8(4.1)    &        0.00(5) \\
\hline\hline
\end{tabular}
\end{table}

A summary of binding parameter errors for H, He, Li, Be, F and Ne are given in table \ref{morse:additional}.
Here, ccECP has the lowest errors when compared to \textit{all} the other ECPs.
In fact, $D_e$ and $r_e$ show smaller errors than UC reflecting the impact of 
taking CC and CV correlations into account.



\begin{table}
\centering
	\caption{Core radii for all ccECPs given in this work. For every channel, radial distance is taken to be the distance from the origin at which angular momentum channel agrees with the bare coulomb potential to within $10^{-5}$ Ha. $R_c$ is the maximum of all channels. Radii are given in Angstrom.}
\label{tab:core_radii}
\begin{tabular}{l|ccccc|cccc}
\hline\hline
 & \multicolumn{5}{c}{\multirow{1}{*}{Local channel included}} &  \multicolumn{4}{c}{\multirow{1}{*}{Nonlocal channel}} \\ 
\hline
Atom & $r_s$ & $r_p$ & $r_d$ & $r_f$ & $R_c$ & $r_s$ & $r_p$ & $r_d$ & $R_c$ \\  
\hline
K  & 0.81 & 0.96 & 0.84 &      & 0.96 & 0.83 & 0.96 &      & 0.96 \\    
Ca & 0.78 & 0.98 & 0.88 &      & 0.98 & 0.87 & 0.99 &      & 0.99 \\    
Ga & 1.95 & 1.82 & 2.78 & 0.59 & 2.78 & 1.95 & 1.82 & 2.78 & 2.78 \\
Ge & 1.45 & 1.60 & 2.34 & 1.51 & 2.34 & 1.48 & 1.58 & 2.34 & 2.34 \\
As & 1.61 & 1.61 & 1.62 & 1.61 & 1.62 & 1.47 & 1.46 & 1.46 & 1.47 \\
Se & 1.18 & 1.33 & 1.64 & 1.08 & 1.64 & 1.18 & 1.33 & 1.64 & 1.64  \\
Br & 1.33 & 1.27 & 1.55 & 1.18 & 1.58 & 1.33 & 1.28 & 1.55 & 1.55  \\
Kr & 1.01 & 1.08 & 1.53 & 0.65 & 1.53 & 1.01 & 1.08 & 1.53 & 1.53  \\
\hline
H        & 0.42 &  & & & 0.42 &  & & &  \\
He       & 0.36 &  & & & 0.36 &  & & &  \\
Li(reg) & 0.59 &  & & & 0.59 &  & & &  \\
Be(reg) & 0.56 &  & & & 0.56 &  & & &  \\
Li & 1.68 & 1.34 & & & 1.68 & 1.68 & & & 1.68  \\
Be & 1.25 & 0.90 & & & 1.25 & 1.25 & & & 1.25  \\
F  & 0.56 & 0.55 & & & 0.56 & 0.55 & & & 0.55  \\
Ne & 0.52 & 0.51 & & & 0.52 & 0.52 & & & 0.52  \\
\hline\hline
\end{tabular}
\end{table}

\subsection{Valence Basis Sets}\label{sec:basis}

For each ccECP pseudoatom, we generated corresponding Gaussian correlation consistent basis sets.
The provided basis sets are cc-pV$n$Z and aug-cc-pV$n$Z with $n=($D, T, Q, 5, 6).
The generation scheme we followed is similar to reference \cite{BFD-2007} which is briefly outlined here.

First, the HF ground state energy is minimized by optimizing a set of primitive exponents.
The number of primitives is chosen in a way that the exact HF energy can be achieved with mHa accuracy or less.
The exponents are optimized in an "even-tempered" way, where only the lowest/highest exponent and the ratio between the exponents are optimized.
Obtained HF orbital coefficients are contracted for every occupied orbital.
For K and Ca, an extra $d$ contraction is also added by exciting an electron to $3d$ orbital.

Then, $m=n-1$ number of correlation consistent primitive exponents are added to every symmetry channel with contraction(s). 
Additional $(n-1, n-2, n-3, ..., 1)$ correlation consistent polarization exponents are added to the next $N=n-1$ higher symmetry channels without a contraction.
For instance, Ge cc-pVQZ will have 1 $s$ and 1 $p$ contraction and additional $(3s,3p,3d,2f,1g)$ correlation consistent primitive exponents.
To optimize the correlation consistent exponents, the configuration-interaction with singles and doubles (CISD) ground state energy of the pseudoatom is minimized.
In cases where there is only 1 valence electron present, the dimer CISD ground state energy is minimized. 
The number and the size of contractions are kept the same for all cardinal numbers $n$ and correlation consistent exponents are also optimized in an even-tempered way as described in reference \cite{corr-basis-scheme}.
Additional diffuse primitive exponents are added by dividing the lowest primitive in each channel by 2.5 to obtain aug-cc-pV$n$Z. 
For K and Ca, $m=n-1$ number of core correlating exponents are added by minimizing the atomic ground state to obtain (aug-)cc-pCVnZ.
For ccECP(reg) basis sets, we use optimized correlation consistent exponents from reference \cite{reg-basis}.
All ccECP basis sets in various code formats can be found at \url{https://pseudopotentiallibrary.org}.

\section{Conclusions}\label{Conclusions}

In this work, we extended our ccECPs to the 3$^{rd}$-row main group elements and provided additional ECPs that complete the ccECP set for all elements from H through Kr. 
The primary departure in the generation of ECPs from previous schemes has been the use of highly accurate many-body theories and methods in the construction.
Another guide in the development was to carry out thorough tests particularly in compressed hydrides and oxides.
These validation tests are imperative since one needs to know the inherent biases in the ECPs in order to be able to assess the corresponding biases in subsequent calculations.

Our ccECPs have substantially improved isospectral and transferable properties compared to previously tabulated ECPs.
We observed that in certain cases such as compressed GaO and GeO, the effective core approximation breaks down. 
While the accuracy of ccECPs in those systems can be improved with more sophisticated parameterizations, the fundamental complication is the missing core-core classical electrostatic interactions which can be improved by considering a smaller core partitioning. 
Previous studies have discussed the complications in ECP construction due to these effects \cite{hay1985ab,schwerdtfeger1995accuracy,leininger1996accuracy} and the significant overlap of the core tails with the valence space \cite{2-ccECP, MgO-ECP}.

On the other hand, we observed that using an ECP is in fact quite beneficial compared to using an uncorrelated core instead. 
Note, for instance, in $4p$ element molecules, UC almost always underbinds considerably and the ccECP is significantly more accurate.
This effect can be seen in spectral discrepancies too, where the ccECP has smaller errors.
The main reason for this can be understood if we recall that the electrons correlate more in an ECP than in UC, i.e, the magnitude of ECP correlation energy is larger than the AE valence-valence counterpart; see for instance reference \cite{Dolg-correlation}.
This is because ECP orbitals have fewer nodes which result in higher probabilities for electrons to correlate.
Therefore, our ccECPs capture part of CV and CC correlations resulting in better molecular properties.
For this reason, the inclusion of total AE correlation energies (CC+CV+VV) is vitally important for proper reproduction of the 
valence properties.

We have shown that the accuracy and transferability of ECPs can be substantially improved by the inclusion of many-body effects.
In fact, we believe that electron correlation has an important contribution to the final accuracy and additional parameters such as core polarization and spin-orbit terms are of secondary importance and can be implemented subsequently. 

One achievement of this work is the demonstration of systems where accurate results are not feasible with conventionally used core approximations.
For example, in LiO and BeO (figure \ref{fig:LiO}, \ref{fig:BeO}) all [He]-core approximations have significant errors and only by including the missing $1s^2$ correlations one is able to recover chemical accuracy.
The improvements represented by  ccECP(reg) can further expand the applicability such as for very high pressures/short bond lengths situations. 
As stated previously, having an upfront knowledge about the ECP biases is important and one can expect these errors to enlarge in solid/bulk calculations.

Another accomplishment of this work including previous papers is that it provides independent tests of previously tabulated ECPs.
For instance, we observed that CRENBL and SBKJC ECPs show very high accuracy in molecules.
SDFSTU pseudopotentials also demonstrate respectable accuracy results in few valence electron atoms.
Independent tests of ccECP have also been carried out recently \cite{ccECP-test}, where the authors carry out selected molecular tests from G2 set using CCSD(T) and fixed-node diffusion Monte Carlo methods.
In both cases, they find that ccECPs have lower MAD in atomization energies from experimental values when compared to other considered ECPs.
Similarly, we believe that ccECPs achieve the best overall consistency in accuracy and we expect that the obtained results for hydrides and oxides will carry over to other systems and cases.

We anticipate that ccECPs will be
also used in 
methods such as DFT, that do not build upon explicitly correlated 
many-body wave functions. Since ccECPs reproduce
the true many-body valence part of the spectrum,
we believe that they should be used as {\em ab initio} valence
Hamiltonians per se, even within such approaches.
In addition, our ccECPs exhibit smooth densities in the core so that one can expect at least some alleviation of DFT biases caused by
the high values and large gradients of the original core densities. However,
one cannot expect the same type of systematic errors from cores in 
all-electronic calculations to automatically transfer into ccECP valence calculations (and procedures similar to nonlinear core corrections \cite{Froyen-Louie-Cohen} might be 
then necessary).
Furthermore, we do not expect much improvement in the 
errors that stem from genuine DFT shortcomings in the valence space (errors in atomizations, gaps, reaction paths, etc). Clearly, further insights and calculations in this direction could be very helpful
and useful.

Future related work is the calculation of accurate atomic ground state energies of ccECPs with various highly precise methods such as CCSDT(Q).
These accurate energies will serve as a benchmark for independent calculations and methods such as fixed-node diffusion Monte Carlo. This will be published in a forthcoming paper.



\section{Supplementary Material}\label{sec:supplementary}
The additional information about ccECPs is included in supplementary material.
Therein are given calculated AE spectra  for each element and also corresponding discrepancies of various core approximations.
AE, UC, and various ECP molecular fit parameters for hydrides, oxides, or dimers are provided.
The ccECPs in semi-local and Kleinman-Bylander projected forms as well as optimized gaussian valence basis sets in various input formats (\textsc{Molpro}, \textsc{GAMESS}, \textsc{NWChem}) can be found at website \cite{ecplibrary}.

Input and output data generated in this work is published in Materials Data Facility \cite{data-facility}. 


\bigskip

{\bf Acknowledgments.}

We are grateful to Paul R. C. Kent for the reading of the manuscript and helpful suggestions.
M. Chandler Bennett contributed to the preparation of this manuscript and helped in formulating the methodologies used for generating the ECPs.

This work has been supported by 
the U.S. Department of Energy, Office of Science, Basic Energy Sciences, Materials Sciences and Engineering Division, as part of the Computational Materials Sciences Program and Center for Predictive Simulation of Functional Materials.

This research used resources of the National Energy Research Scientific Computing Center (NERSC), a U.S. Department of Energy Office of Science User Facility operated under Contract No. DE-AC02-05CH11231.

This paper describes objective technical results and analysis. Any subjective views or opinions that might be expressed in the paper do not necessarily represent the views of the U.S. Department of Energy or the United States Government.

Sandia National Laboratories is a multimission laboratory managed and operated by National Technology \& Engineering Solutions of Sandia, LLC, a wholly owned subsidiary of Honeywell International Inc., for the U.S. Department of Energy's National Nuclear Security Administration under contract DE-NA0003525. 


\bibliography{main.bib}

\end{document}